\documentclass[preprint2]{aastex62}
\usepackage{amsmath,amstext}
\usepackage[T1]{fontenc}
\usepackage{apjfonts} 
\usepackage[figure,figure*]{hypcap}

\newcommand{\cutt}[1]{\textcolor{blue}{}}

\newcommand{\Ms}{{\ensuremath{{M}_{\odot} }}}
\newcommand{\Zs}{\ensuremath{Z_\odot}}
\newcommand{\Cx}{{\ensuremath{^{12}\mathrm{C}}}}
\newcommand{\Ni}{{\ensuremath{^{56}\mathrm{Ni}}}}
\newcommand{\Fe}{{\ensuremath{^{56}\mathrm{Fe}}}}
\newcommand{\Co}{{\ensuremath{^{56}\mathrm{Co}}}}
\newcommand{\He}{{\ensuremath{^{4} \mathrm{He}}}}
\newcommand{\Hy}{{\ensuremath{^{1} \mathrm{H}}}}

\newcommand{\Ox}{{\ensuremath{^{16}\mathrm{O}}}}
\newcommand{\Ti}{{\ensuremath{^{44}\mathrm{Ti}}}}
\newcommand{\Si}{{\ensuremath{^{28}\mathrm{Si}}}}
\newcommand{\Ne}{{\ensuremath{^{20}\mathrm{Ne}}}}
\newcommand{\Cr}{{\ensuremath{^{48}\mathrm{Cr}}}}
\newcommand{\Ca}{{\ensuremath{^{40}\mathrm{Ca}}}}
\newcommand{\Ar}{{\ensuremath{^{36}\mathrm{Ar}}}}

\newcommand{\Mg}{{\ensuremath{^{24}\mathrm{Mg}}}}
\newcommand{\gcc}{\ensuremath{\mathrm{g}\,\mathrm{cm}^{-3}}}

\newcommand{\FIGFF}[2]{{\ref{fig:#2}{#1}}}

\newcommand{\FIG}[2]{{Fig.~\FIGFF{#1}{#2}}}
\newcommand{\Fig}[1]{{\FIG{}{#1}}}
\newcommand{\lFig}[1]{{\label{fig:#1}}}

\newcommand{\CASTRO}{\texttt{CASTRO}}

\newcommand{\KEPLER}{\texttt{KEPLER}}

\newcommand{\STELLA}{\texttt{STELLA}}

\newcommand{\ken}[1]{\textcolor{black}{ #1}}

\shorttitle{\Ni\ Heating in PI SNe}
\shortauthors{Chen, Woosley, \& Whalen}

\begin{document}

\title{Gas Dynamics of the Nickel-56 Decay Heating in Pair-Instability Supernovae}

\author{Ke-Jung Chen}
\affiliation{Institute of Astronomy and Astrophysics, Academia Sinica, Taipei 10617, Taiwan} 
\affiliation{Division of Theoretical Astronomy, National Astronomical Observatory of Japan, Tokyo 181-8588, Japan} 

\author{S. E. Woosley}
\affiliation{Department of Astronomy and Astrophysics,
	University of California at Santa Cruz, Santa Cruz, CA 95060, USA}
\author{Daniel J. Whalen}
\affiliation{Institute of Cosmology and Gravitation, University of Portsmouth, Portsmouth PO1 
	3FX, UK}
\affiliation{Ida Pfeifer Professor, Department of Astrophysics, University of Vienna, Tuerkenschanzstrasse 17, 1180, Vienna, Austria}

\correspondingauthor{Ke-Jung Chen}
\email{kjchen@asiaa.sinica.edu.tw}

\begin{abstract} 

Very massive $140-260$ \Ms\ stars can die as highly-energetic pair-instability supernovae (PI SNe) with energies of up to 100 times those of core-collapse SNe that can completely destroy the star, leaving no compact remnant behind.  These explosions can synthesize $0.1-30$ \Ms\ of radioactive \Ni, which can cause them to rebrighten at later times when photons due to \Ni\ decay diffuse out of the ejecta.  However, heat from the decay of such large masses of \Ni\ could also drive important dynamical effects deep in the ejecta that are capable of mixing elements and affecting the observational signatures of these events.  We have now investigated the dynamical effect of \Ni\ heating on PI SN ejecta with high-resolution two-dimensional hydrodynamic simulations performed with the \CASTRO\ code.  We find that expansion of the hot \Ni\ bubble forms a shell at the base of the silicon layer of the ejecta $\sim$ 200 days after the explosion but that no hydrodynamical instabilities develop that would mix \Ni\ with the \Si/\Ox-rich ejecta.  However, while the dynamical effects of \Ni\ heating may be weak they could affect the observational signatures of some PI SNe by diverting decay energy into internal expansion of the ejecta at the expense of rebrightening at later times.  

\end{abstract}


\section{Introduction}

Observations by \citet{humphrey1979}, \citet{davidson1997}, and \citet{r136} indicate that stars with masses $>$ 100 \Ms\ can form in the local universe.  Cosmological simulations also suggest that the initial mass function (IMF) of primordial (or Pop III) stars is top-heavy and that many of them would have died with masses above 100 \Ms\ \citep[e.g.,][]{hir14}.  Very massive stars (VMS; $140-260$ \Ms) are thought to explode as pair-instability supernovae \citep[PI SNe;][]{woosley2002}.  The original idea of the pair instability was introduced by \citet{bark1967} and \citet{rs67} and further developed by \citet{ober1983}, \citet{glatzel1985}, \citet{stringfellow1988}, \citet{heger2002} and \citet{heger2010}.  When the core of a VMS evolves to temperatures above 10$^9$ K, thermal photons in the tail of their Maxwellian distribution (h$\nu \ge$ 1 Mev) begin to freeze out as electron - positron ($e^-$--$e^+$) pairs through collisions with nuclei.  Pair-production occurs at the expense of thermal pressure support of the core against gravity and it begins to contract and become hotter.  Rising temperatures and densities in the core eventually ignite explosive oxygen and silicon burning that can completely disrupt the star.  PI SNe can produce $10^{52}-10^{53}$ erg of energy and $0.1- 30$ \Ms\ of \Ni\ and may be the most energetic thermonuclear explosions in the universe.  Unlike core-collapse (CC) SNe, whose central engines are not fully understood, PI SN explosions are much simpler and are emergent features of stellar evolution models.

Stellar evolution models suggest that the progenitors of PI SNe can die with a variety of structures, ranging from red supergiants (RSGs) to blue supergiants (BSGs) whose radii differ by a factor of 100.  \citet{Cha15} recently found that stellar rotation can even cause PI SN progenitors to lose their hydrogen envelopes and die as bare helium cores.  They also found that rotation can shift the lower mass limit of PI SN progenitors from 140 \Ms\ down to to 85 \Ms\ (\Ni\ production can drop sharply in such events; \citealt{chen2015}).  It was originally thought that the shells of elements built up in the interior of the star over its life would be severely disrupted and mixed by the passage of the PI SN shock through them prior to breakout from the surface.  However, multidimensional models have since shown that the star expands homologously during the explosion, with little mixing between shells \citep{candace2011,chen2011,chen14a,Gil17}.  

PI SN light curves and spectra depend heavily on the structure of the star at death and \Ni\ production and internal mixing during the explosion \citep[e.g.,][]{kasen2011,det12,wet12b,wet13d,smidt14a,Koz14,Koz15,Maz19}.  They are generally characterized by an intense, brief radiation pulse when the shock breaks out of the star followed by a decline in luminosity as the fireball expands and cools.  The luminosity can then rise again after a few weeks to months when photons due to the decay of \Ni\ begin to diffuse out of the ejecta.  This rebrightening phase can last from weeks to months depending on \Ni\ mass and the structure of the star.  

PI SNe could be the ultimate cosmic lighthouses because they can be detected in the near infrared (NIR) at cosmic dawn at $z \sim$ 25 by the {\it James Webb Space Telescope} ({\it JWST}) and at later epochs by the {\it Nancy Grace Roman Space Telescope} and the next generation of extremely large telescopes \citep{wet12a}.  They therefore could probe the masses of the first generation of stars \citep{hum12,chat2012a,pan12a,wet12a,mw12,ds13,ds14,mes13a,Har18} and the origins of  extremely metal-poor stars \citep{candace2010,ish18,Tak18}.  RSG PI SNe have the strongest NIR signals at high redshift while those of BSGs and stripped helium cores are much weaker \citep{wet13e,smidt13a,smidt14a}.  Several PI SN candidates have now been found \citep[e.g.,][]{2007bi,cooke12,pisn19,pisn20}.

A unique aspect of PI SNe are the large masses of \Ni\ they produce, and the effects of its decay on their light curves, spectra and ejecta dynamics are not fully understood.  5 \Ms\ of \Ni\ releases $\sim 1\times10^{51}$ erg (about the energy of a CC SN) as $\gamma$-rays.  They downscatter in energy as they diffuse out of the ejecta on timescales that depend on the mass between the \Ni\ and the surface of the star, and thus its structure at death.  Many light curve and spectrum models take nearly all of these photons to escape and cause the SN to rebrighten at later times when in reality only some do because the remainder heat material deep in the ejecta and create a hot bubble that expands and does $PdV$ work on its surroundings.  In principle, the expansion of this bubble could trigger the formation of hydrodynamical instabilities and mix \Ni\ with other elements.  Consequently, \Ni\ heating in PI SNe could reduce rebrightening and change the order in which lines appear in the spectra at later times.

These issues have never been resolved because no multidimensional explosion model has ever been run for long enough times to determine the dynamics of the hot bubble (> 100 days).  We have now evolved PI SNe out to 300 days in two-dimensional (2D) simulations to evaluate the effects \Ni\ heating on the dynamics of the ejecta.  Our numerical method and explosion models are described in Section 2.  The dynamics of PI SNe at later times (and of the \Ni\ bubble in particular) are examined in Section 3.  We discuss the implications \Ni\ heating in Section 4 and conclude in Section 5.

\section{Numerical Method}

Our three PI SN progenitor models were first exploded  in 1D in \KEPLER.  The blast profiles were then mapped onto a 2D grid in \CASTRO\ and evolved out to 300 days.

\subsection{\KEPLER\ Models}

Rotating VMS can produce bare He cores, as discussed earlier, while mixing in the cores of even Pop III stars can create RSGs with radii 100 times greater than those of the main-sequence star \citep[e.g.,][]{meakin2007,arnett2009,woodward2013}.  After the main sequence, carbon and oxygen in the central convective cores of massive stars become closer to the hydrogen-burning shell.  If any convective overshooting or other convective boundary mixing occurs, these heavier elements can be mixed into the hydrogen layer, dramatically increasing energy production in the core and expanding the star into an RSG.  We therefore chose three models that bracket the likely range of structures of PI SN progenitors: 105 \Ms\ and 110 \Ms\ He cores (models He105 and He110) and a $Z =$ 10$^{-4}$ \Zs\ 225 \Ms\ RSG that has retained its hydrogen envelope (model U225).  These stars explode with energies of 48.3 B, 55.3 B and 46.6 B (1 B $=$ 10$^{51}$ erg) and produce 8.53 \Ms, 13.13 \Ms\ and 16.52 \Ms\ of \Ni, respectively.  The physical properties of all three stars and their explosions are listed in Table \ref{tbl:models}. 

We evolve all three stars from the onset of the PI through core contraction, explosion and the end of all nuclear burning in the \KEPLER\ stellar evolution code \citep{kepler}.  Our simulations were initialized with profiles from \citet{heger2010} and evolved until there was no further change in the energy or composition of the ejecta, $\sim 10^3 -10^4$ sec after the explosion.  At this time the forward shock in He105 and He110 has just broken through the surface of the star while it has only entered the hydrogen envelope in U225.  We perform the explosion in 1D in \KEPLER\ because it naturally emerges from the stellar evolution calculation and we can follow nucleosynthesis in much more detail than would be practical in 2D or 3D.  

\ken{There are also no significant departures from spherical symmetry in the ejecta over the short times we evolve the explosion in \KEPLER. \cite{chen2011,chen14a} found that explosive burning did drive the formation of some instabilities in the O shell but they could not grow to large amplitudes because burning ended after a few tens of seconds.  They are much smaller than those in CC SNe.  We approximate these structures in our 2D \CASTRO\ models by seeding the grid with random velocity perturbations of the order of $\sim$ 1\% of the local sound speed}.

\begin{deluxetable*}{lcccccc}
	\tablecaption{1D \KEPLER\ PI SN Models}
	\tablehead{
		\colhead{Model} &
		\colhead{Stellar mass} &
		\colhead{Stellar radius} &
		\colhead{He core mass} &
		\colhead{Explosion energy} &
		\colhead{\Ni\ production} &
		\colhead{\Ni\ decay energy}  
		\\
		\colhead{} &
		\colhead{($\Ms$)} &
		\colhead{($10^{12}$ cm)} &
		\colhead{($\Ms$)}    &
		\colhead{($10^{51}$ erg)} &
		\colhead{($\Ms$)}      & 
		\colhead{($10^{51}$ erg)} 
	}
	\startdata
	He105   &  105 &   1.42  &  105  & 48.3 &    8.53  & 1.58\\
	He110   &  110  &   1.43  &   110  &  55.3  &  13.13   & 2.47\\
	U225   &  225  &   334   &  104   &  46.6  & 16.52 	& 3.07\\
	\enddata
	
	\tablecomments{The decay energy of 1 \Ms\  \Ni{} $\to$ \Co{} $\to$ \Fe{}  decay is $ 1.86\times10^{50}$ erg. }
	\vspace{0.2 in}
	\label{tbl:models}
\end{deluxetable*}  

We show elemental abundances at the end of all three \KEPLER\ runs in \Fig{fig:ic1}, which show the onionlike structure of elements in the ejecta.  No \Ni\ overlaps with the \Cx\ or \Ox\ layers. Corresponding density and velocity profiles are shown in \Fig{fig:ic2}.  The velocity of the forward shock in He105 and He110 is $\sim 2\times10^9$ cm sec$^{-1}$.  The helium stars have become unbound and their central densities have dropped to $\sim 200$ \gcc.  In the U225 model the shock has just entered the hydrogen envelope and its velocity is also $\sim 2\times10^9$ cm s$^{-1}$. Its central density is  $\sim 1$ \gcc.  The absence of any contact discontinuities in these profiles indicates that no fluid instabilities would have formed in these three explosions by these times in multidimensional simulations.  Therefore, if any mixing occurs it is due to the propagation of the shock or the expansion of the \Ni\ bubble at later times. 

\begin{figure}
	\centering
	\includegraphics[width=.9\columnwidth]{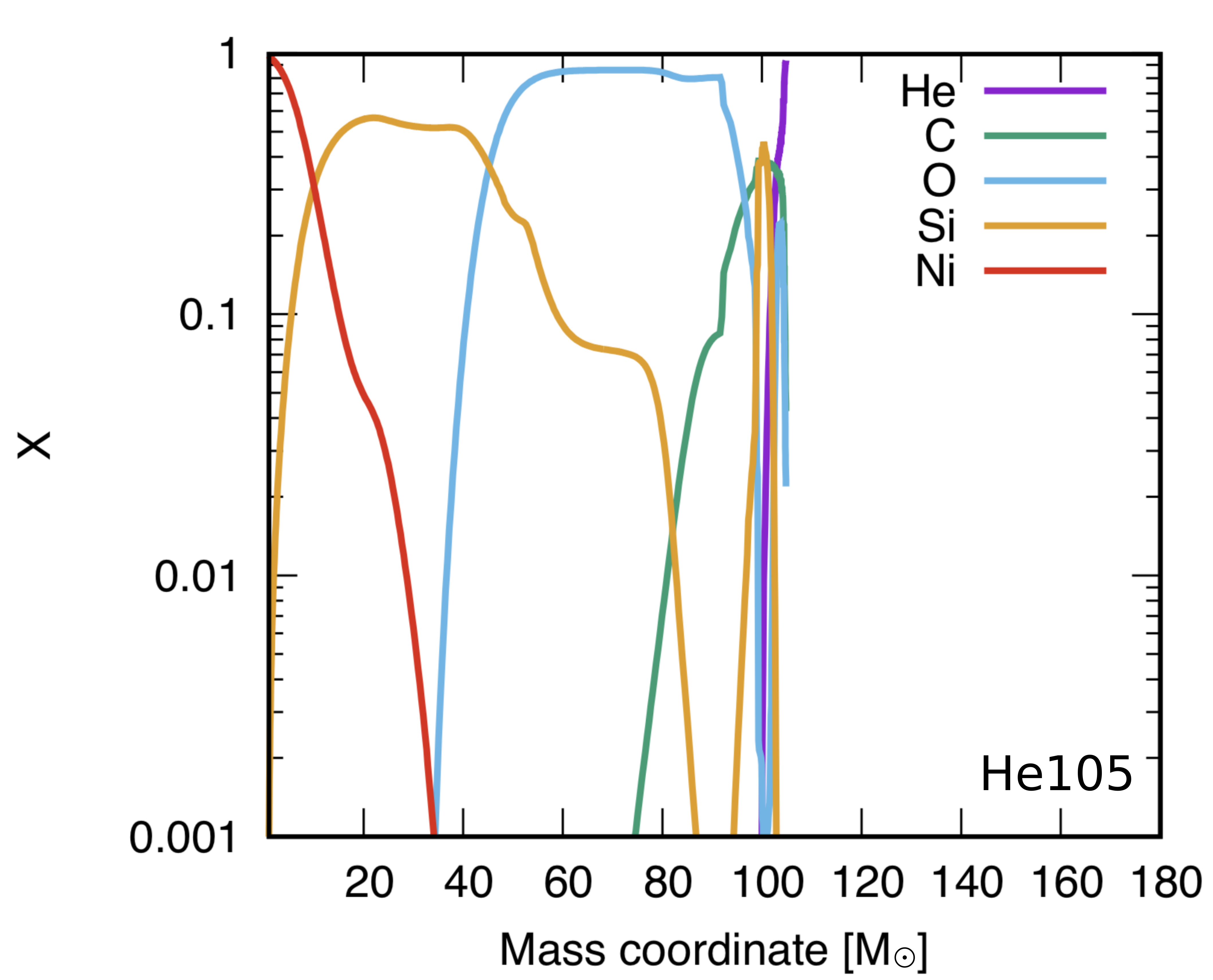}		 	
	\includegraphics[width=.9\columnwidth]{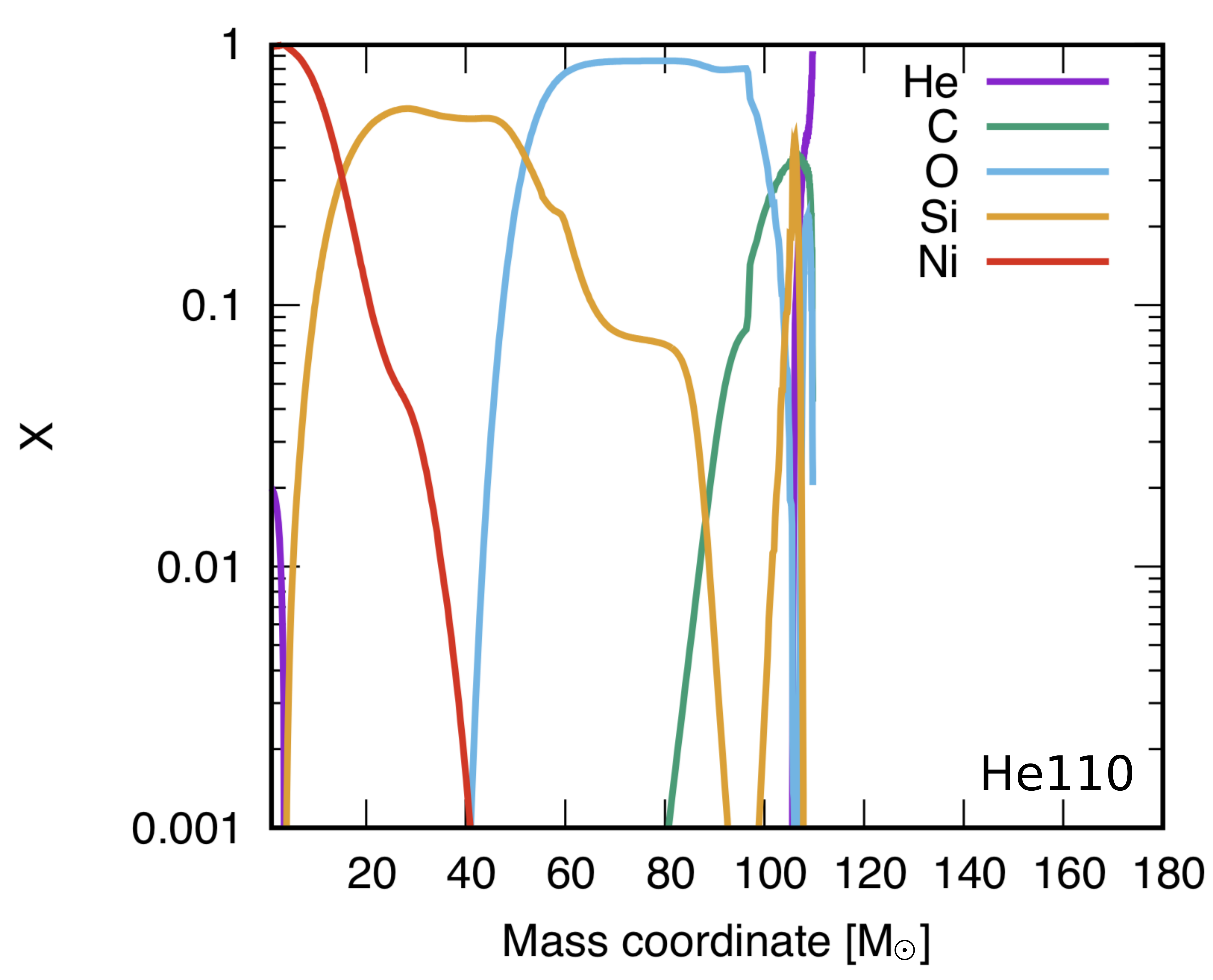}
	\includegraphics[width=.9\columnwidth]{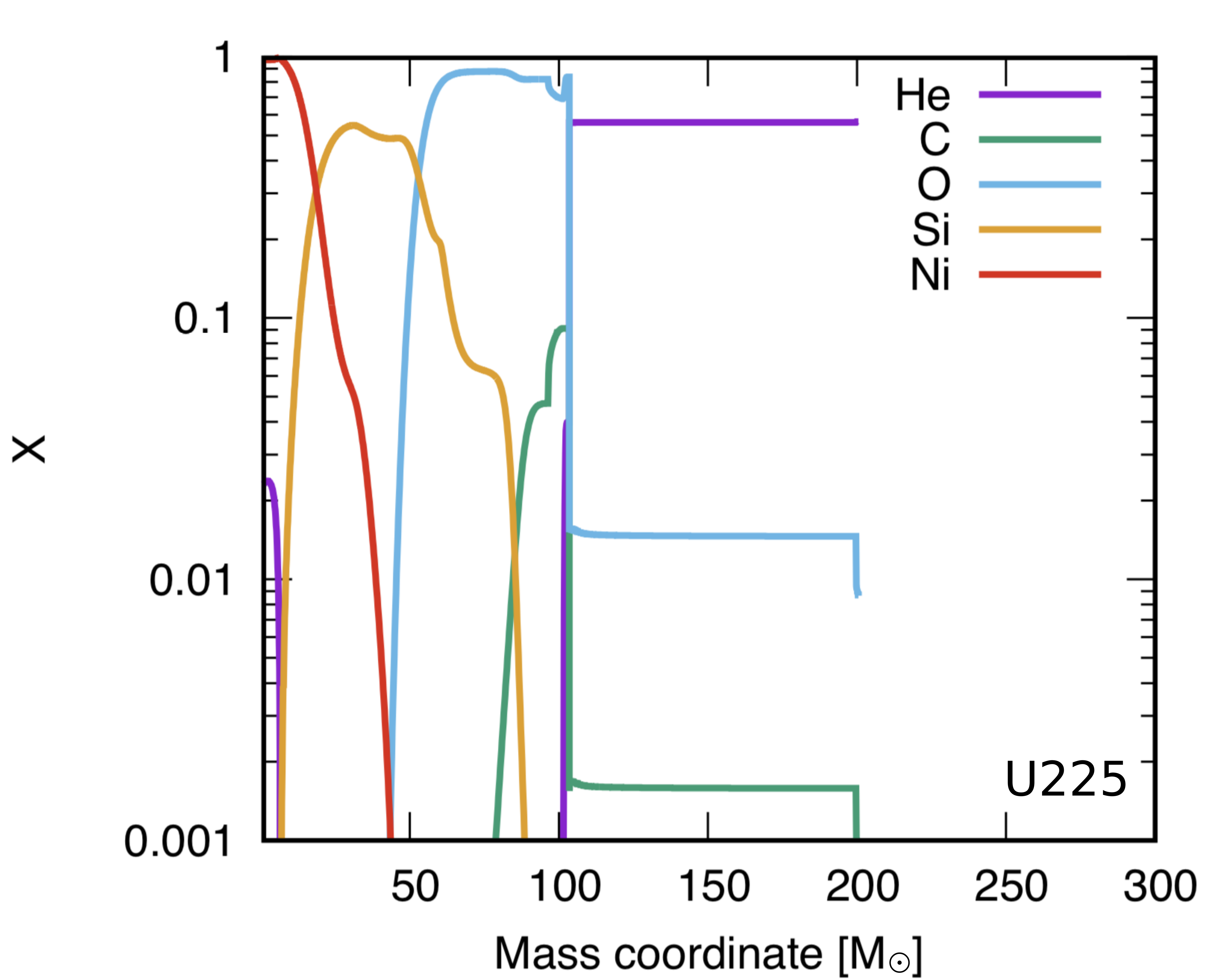}		 		
	\caption{\lFig{fig:ic1} Final 1D \KEPLER\ He, C, O, Si and Ni mass fractions, which show the onion-like structure of elemental shells that have built up in the stars over their lives.  In none of the stars does Ni overlap with the C or O shells.  Note that convective mixing has dredged C and O up into the He and H shells in U225.}
\end{figure}

%
%
%
%
\begin{figure}
	\centering
	\includegraphics[width=.72\columnwidth]{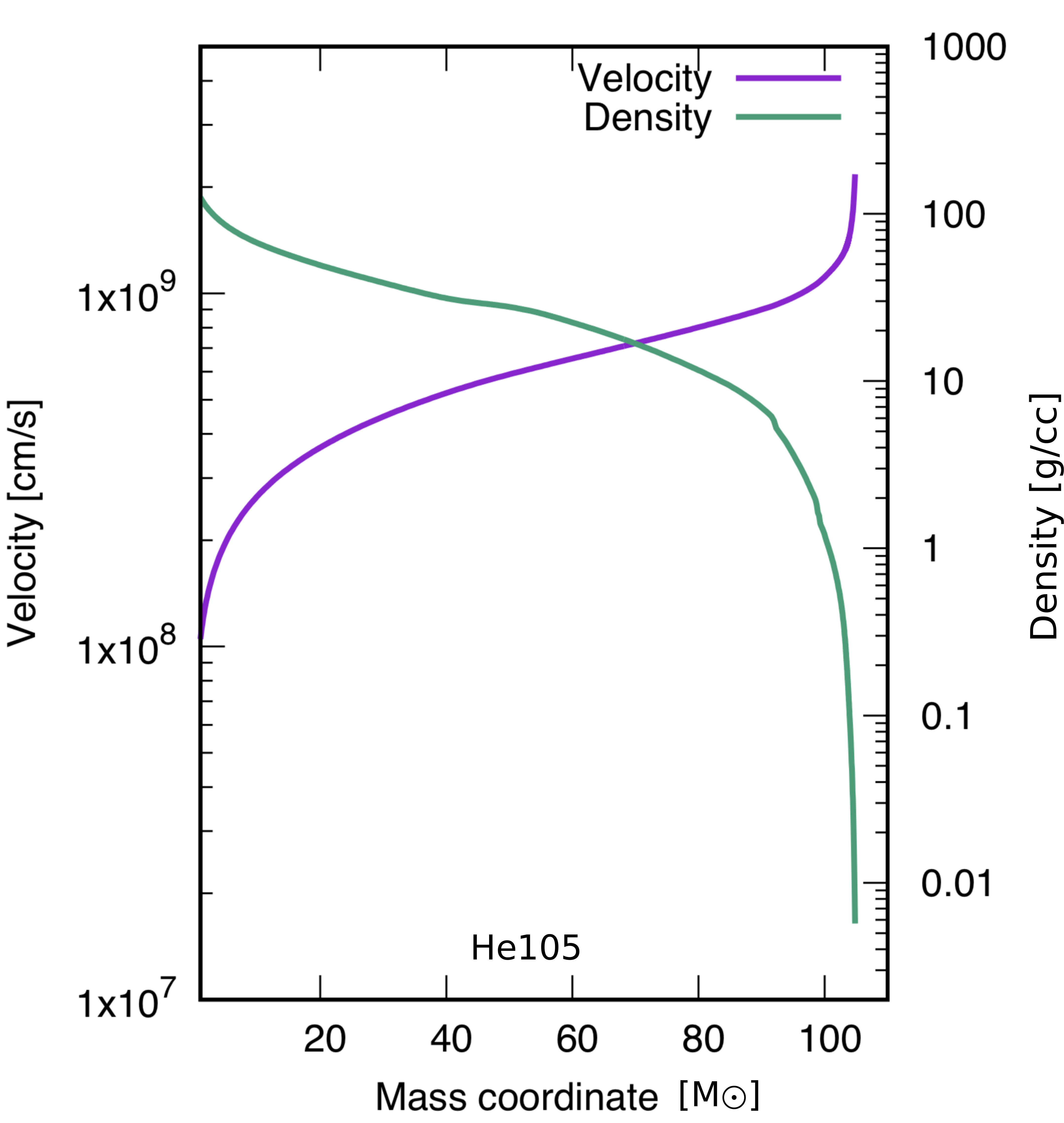}		 	
	\includegraphics[width=.72\columnwidth]{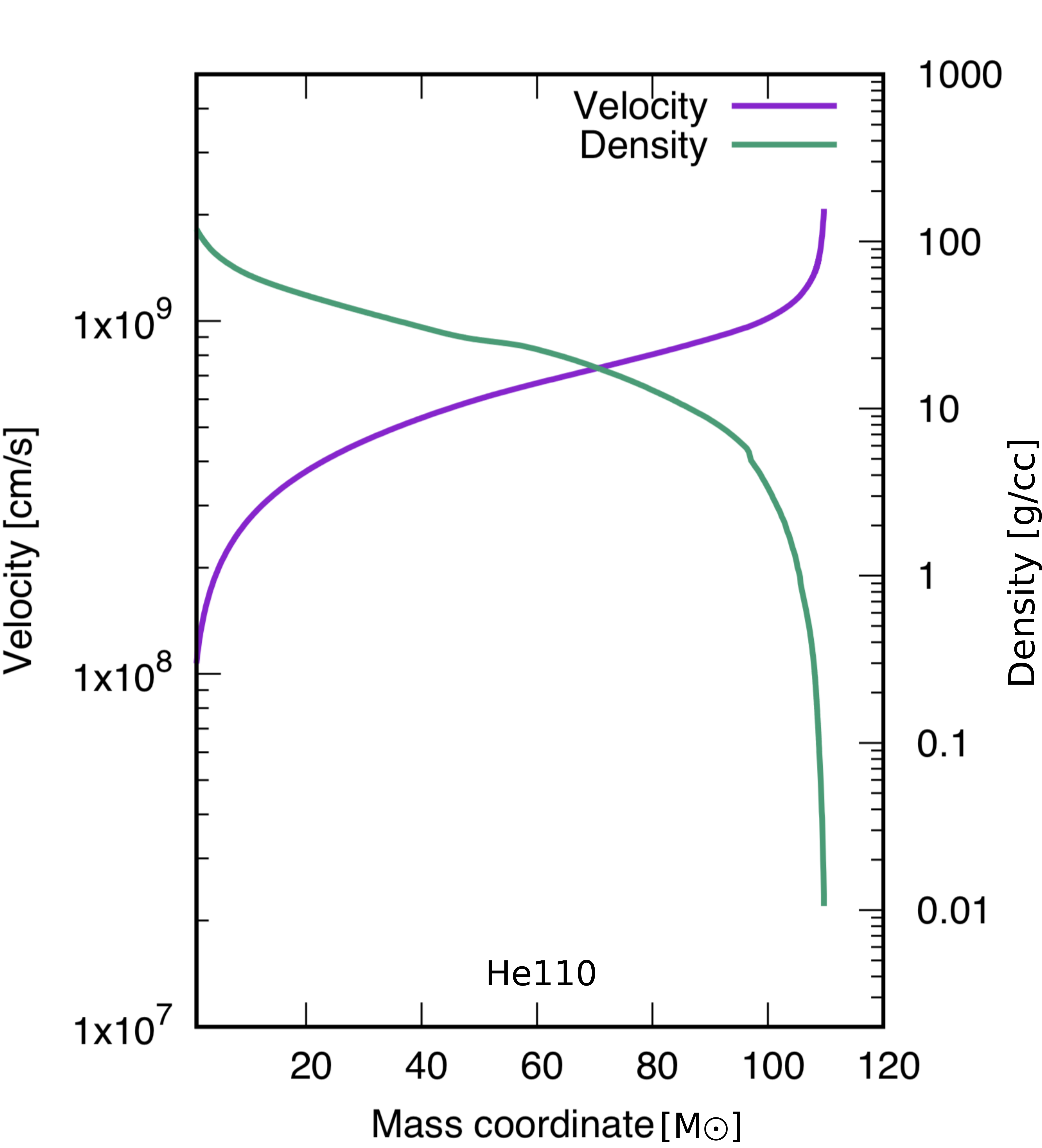}
	\includegraphics[width=.72\columnwidth]{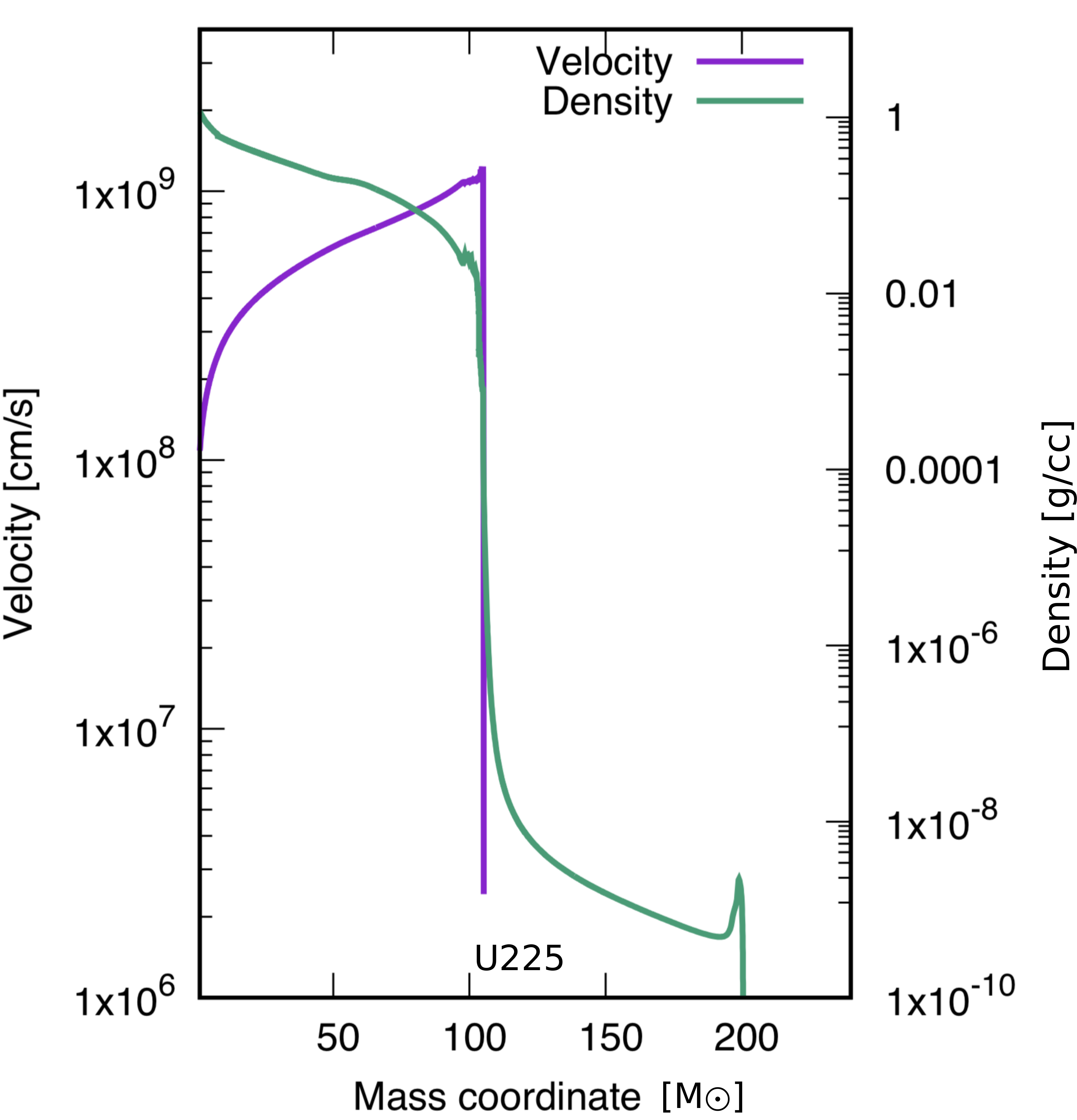}		 		
	\caption{\lFig{fig:ic2} Final 1D \KEPLER\ density and velocity profiles.  The forward shock has just broken through the stellar surface in the He105 and He110 models and entered the hydrogen envelope in the U225 model.}
\end{figure}
%
 
%
%
 
\begin{figure}
	\centering
	\includegraphics[width=.7\columnwidth]{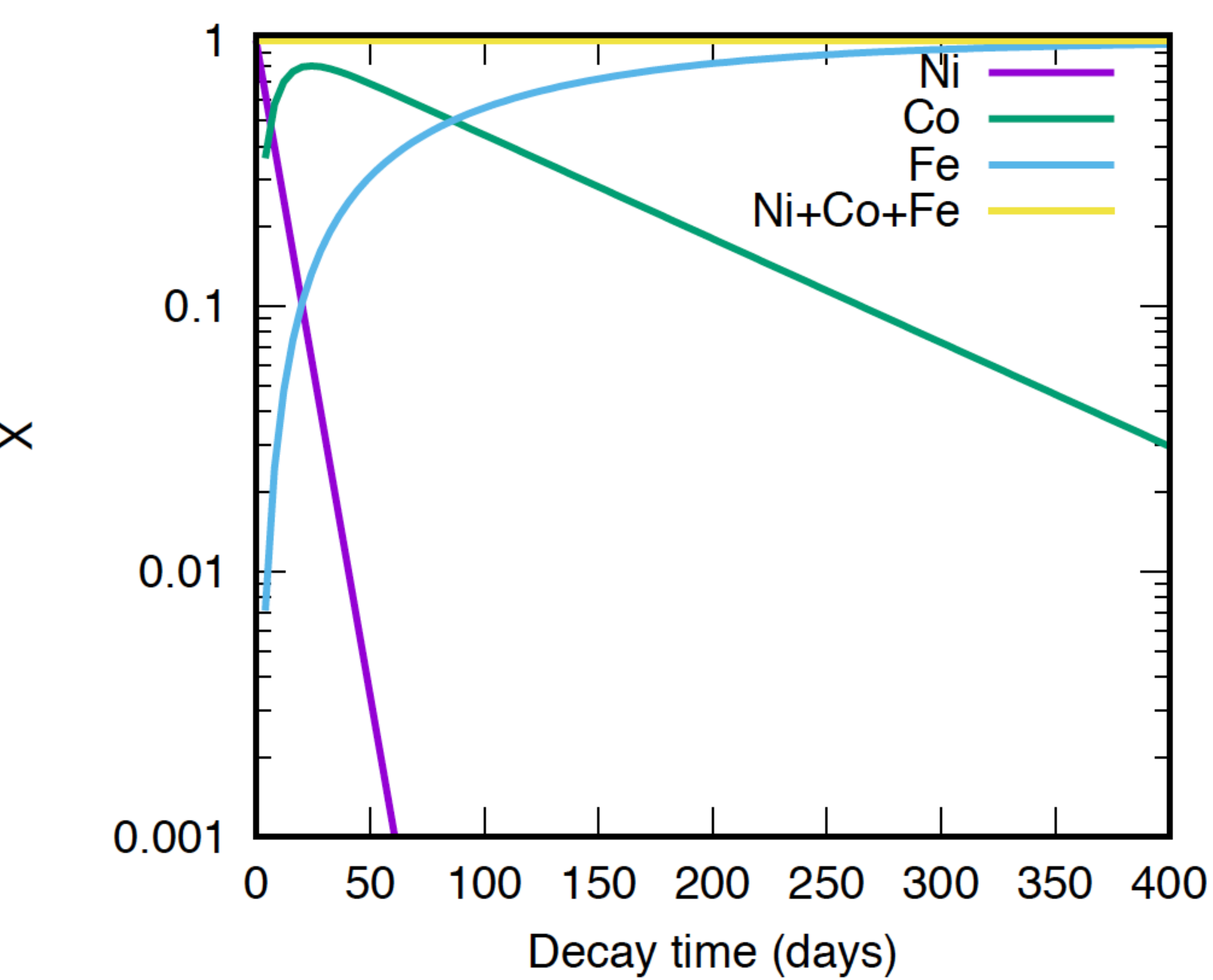}		 	
	\includegraphics[width=.7\columnwidth]{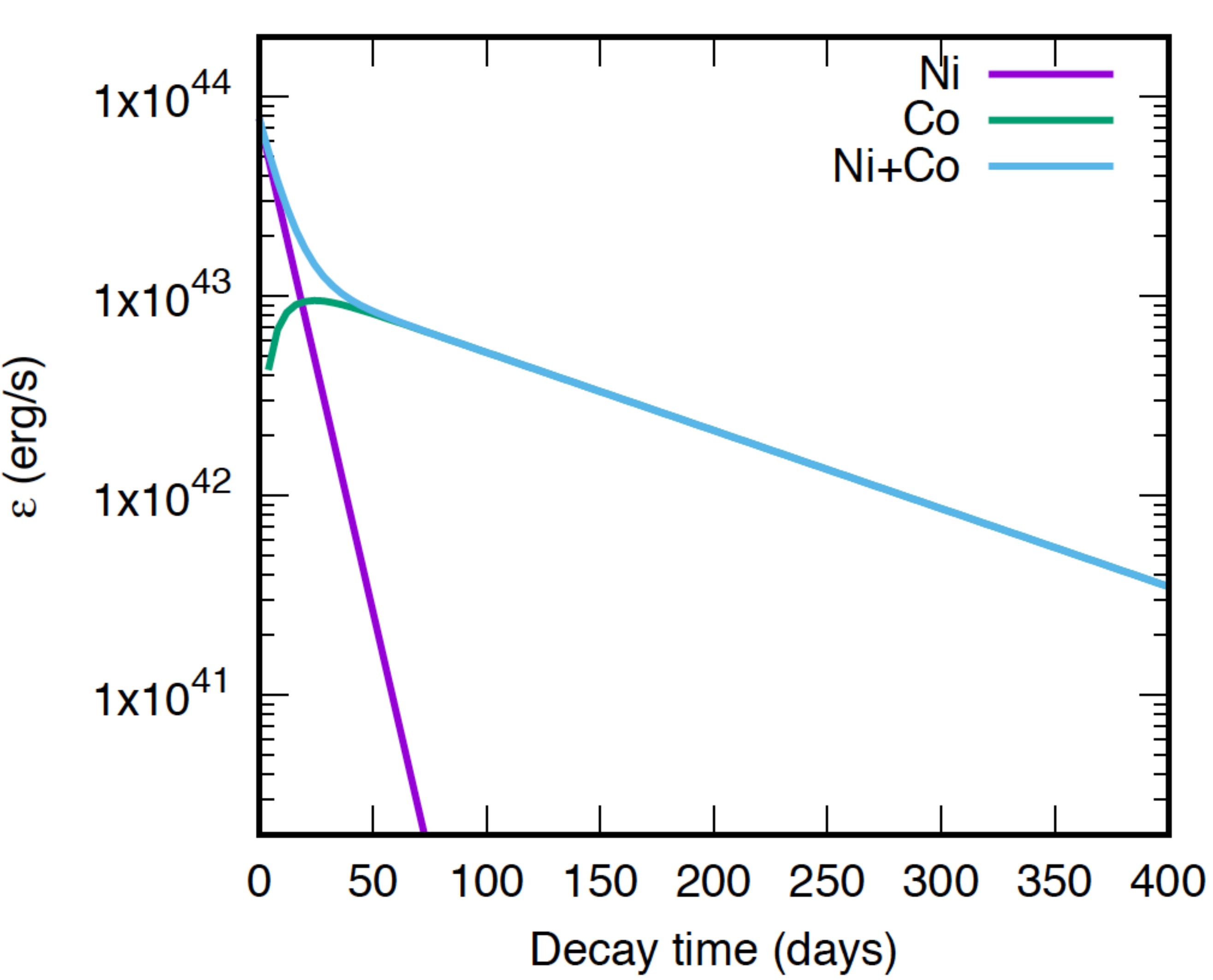}
	\includegraphics[width=.7\columnwidth]{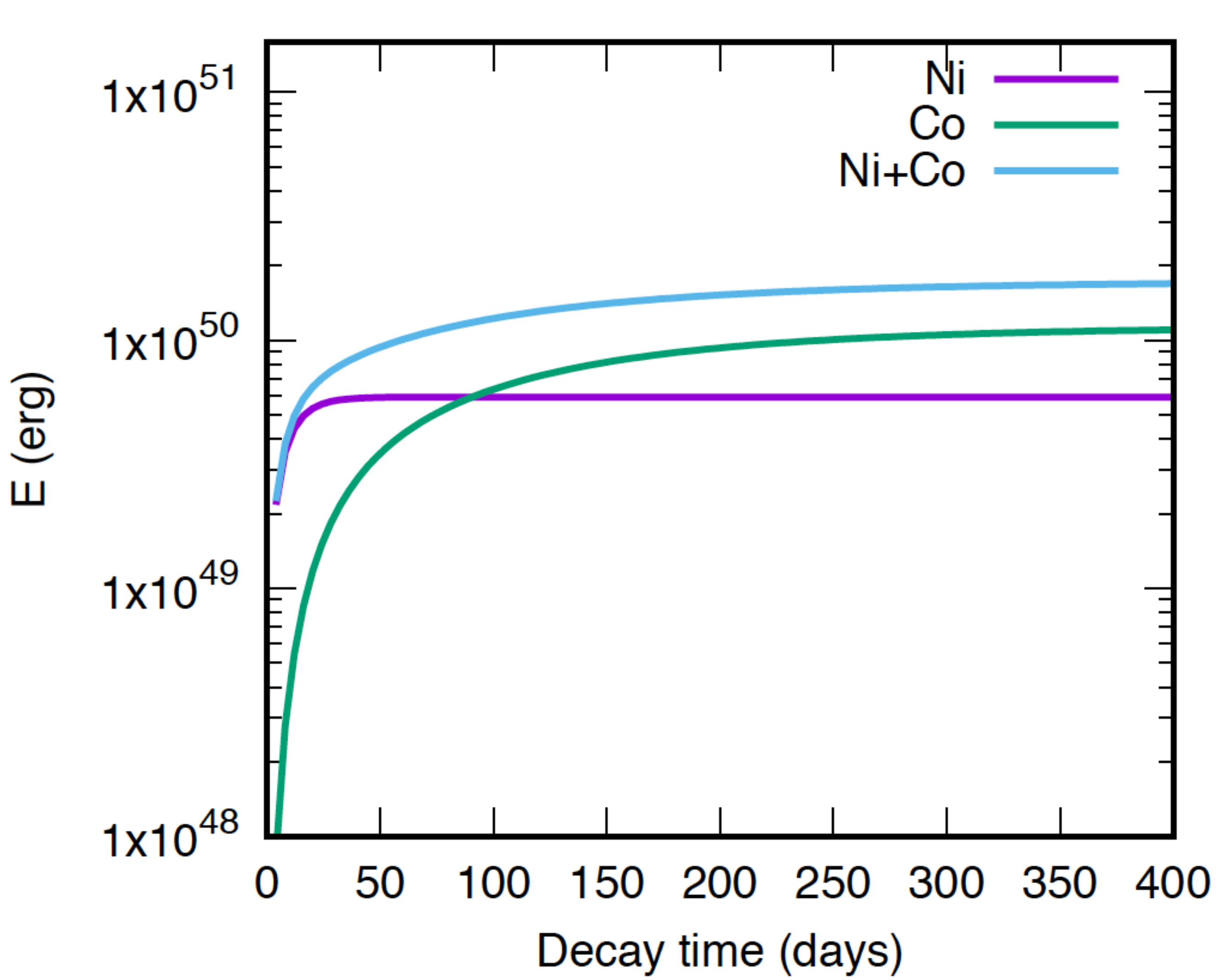}		 		
	\caption{\lFig{fig:nidecay} Decay of 1 \Ms\ of \Ni.  Top panel:  \Ni, \Co, and \Fe\ mass fractions. About 90\% of the \Ni\ decays into \Co\ within 20 days.  Middle panel:  energy generation rates due to \Ni\ and \Co\ decay.  In the first 50 days they can reach $10^{43}$ erg s$^{-1}$.  If all this energy escapes the ejecta as radiation it produces a bright UV/optical transient at later times.  Energy generation at later times (> 50 days) is dominated by \Co\ decay.  Bottom panel:  total energy released by radioactive decay.  The total energy released by $\Ni\rightarrow\Co$ and $\Co\rightarrow \Fe$ decay is $5.92 \times 10^{49}$ erg and $1.28 \times 10^{50}$ erg, respectively.}
\end{figure}

\subsection{2D \CASTRO\ Simulations}

We then port our 1D \KEPLER\ blast profiles onto a 2D cylindrical grid in \CASTRO\ with the conservative mapping scheme developed by \citet{chen2013}.  This method preserves the mass, energy and momentum of the ejecta on the new grid over a large range of spatial scales from features in the O and Si shells at $\sim$ 10$^9$ cm to the radius of the star itself, which is four to six magnitudes larger.  \CASTRO\ is a multidimensional adaptive mesh refinement (AMR) hydrodynamics code for astrophysical simulations \citep{ann2010, zhang2011}.  \CASTRO\ uses an unsplit piecewise parabolic method (PPM) hydro scheme \citep{woodward1984} with multispecies advection and has several different equations of state (EOS). 

We advect the thirteen species that constitute the PI SN ejecta:  \Hy, \He, \Cx, \Ox, \Ne, \Mg, \Si, \Ar, \Ca, \Ti, \Cr, \Fe, and \Ni.  We use the Helmholtz EOS \citep{timmes2000}, which includes contributions by both degenerate and non-degenerate relativistic and non-relativistic electrons, positron-electron pairs, ions, and radiation, during the early phase of the explosion and switch to an ideal gas EOS later when densities fall below 10$^{-12}$ \gcc.  The gravity solver applies the monopole approximation by constructing a spherically symmetric gravitational potential from the radial average of the density and then calculating the corresponding gravitational force everywhere in the AMR hierarchy.  It is a reasonable approximation because departures from spherical symmetry in the ejecta are small.

Our \CASTRO\ root grid is $1.875 \times 10^{14}$ cm in $r$ and $z$ with a resolution of 256$^2$.  We center eight nested grids on the star for a maximum spatial resolution of 256 $\times$ 2$^8$, or $2.86 \times 10^9$ cm.  This zone size was found by \citet{chen14a} to be sufficient to resolve the onset of fluid instabilities driven by explosive Si and O burning in multidimensional PI SN simulations.  Up to eight levels of AMR refinement are also allowed during the initial mapping of the \KEPLER\ profiles into \CASTRO\ and the run thereafter.  Refinement is done on gradients in density, velocity, and pressure.  We use both static nested grids and AMR because the nested grids ensure that the \Ni-rich region always receives the highest resolution while the AMR properly resolves important flows that are outside the most deeply nested grids.  Reflecting and outflow boundary conditions are applied to the simulation box on the inner and outer boundaries in $r$ and $z$, respectively, and we simulate one octant of the explosion.

Because we evolve the explosion for 300 days the SN shock soon reaches the outer boundaries of the original simulation domain.  \ken{When the shock reaches a grid boundary we quadruple the size of the mesh while holding the number of mesh points constant and conservatively map the flows onto this new grid using the approach by \citet{chen2013}.  In our runs the original box is quadrupled three times to a final size of $\sim 1.2 \times10^{16}$ cm, which is $\sim$ 100 and 10,000 times larger than the radii of the RSG and helium stars, respectively (note that the original mesh fully enclosed the helium stars but not the RSG)}.  

We surround all three stars with a circumstellar medium (CSM) density profile of the form $\rho =10^{-4}\rho_{\mathrm{s}} (\frac{r}{r_{\mathrm{s}}})^{-2}$, where $\rho_{\mathrm{s}}$ and $r_{\mathrm{s}}$ are the surface density and radius of the star.  This is done to prevent gas densities from becoming negative after the shock crashes out of the star.  The total masses of these envelopes out to the final, largest mesh boundaries are $\sim 4\times10^{-5}$ \Ms\ for the helium stars and $\sim 0.05$ \Ms\ for the RSG, which are negligible in comparison to the masses of the stars themselves.  We chose diffuse envelopes to prevent the formation of reverse shocks that could drive the growth of Rayleigh-Taylor (RT) instabilities at later times as the SN shock plows up the ambient medium.  Although the CSM of actual stars like those in our study are not known, the ones we chose are consistent with the weak winds and mild mass loss that are usually associated with low-metallicity stars ($\rho \propto r^{-2}$).

\subsection{\Ni\ Decay}

\Ni\ is synthesized during PI SNe at the center of the star by explosive Si burning.  It decays with a half-life of 6.1 days to \Co{}, which then decays with a half-life of 77.1 days to \Fe{}.  The $\gamma$-rays heat the ejecta as they downscatter in energy through it.  We do not transport $\gamma$-rays in \CASTRO\ so we deposit their energy locally as internal energy.  This approximation holds as long as the surrounding ejecta is optically thick to $\gamma$-rays.  The energy generation rate per unit volume for \Ni\ decay, $\dot{\epsilon}_{\Ni}$, is:
\begin{equation}
\dot{\epsilon}_{\Ni}(t) =\lambda_{\Ni}\rho X_{\Ni}Q_\Ni\ e^{-\lambda_{\Ni}t}, 
\end{equation}
where $X_{\Ni}$ is the initial \Ni\ mass fraction, $\rho$ is the gas density, $\lambda_{\Ni} = 1.315\times10^{-6}$ s$^{-1}$ is the initial \Ni\ decay rate and $Q_\Ni \sim 2.96 \times 10^{16}$ erg g$^{-1}$ is the energy released per gram of \Ni\ \citep{nad94}.  The \Co\ mass fraction, $X_{\Co}(t)$, is a function of $X_{\Ni}$:
\begin{equation}
	X_{\Co}(t) =
	\frac{\lambda_{\Ni}}{\lambda_{\Ni}-\lambda_{\Co}}
	X_{\Ni} (e^{-\lambda_{\Co}t}-e^{-\lambda_{\Ni}t}),
\end{equation}
so the energy generation rate due to the decay of \Co, $\dot{\epsilon}_{\Co}$, is:
\begin{equation}
	\dot{\epsilon}_{\Co}(t) =\frac{\lambda_{\Co}\lambda_{\Ni}}{\lambda_{\Ni}-\lambda_{\Co}}
	\rho X_{\Ni} Q_\Co(e^{-\lambda_{\Co}t}-e^{-\lambda_{\Ni}t}), 
\end{equation}
where $\lambda_{\Co} = 1.042 \times10^{-7}$ sec$^{-1}$ is the initial \Co\ decay rate and $Q_\Co \sim 6.4 \times 10^{16}$ erg g$^{-1}$ is the energy released per gram of \Co\ \citep{nad94}. 

\begin{figure*}[h]
	\centering
	\includegraphics[width=\textwidth]{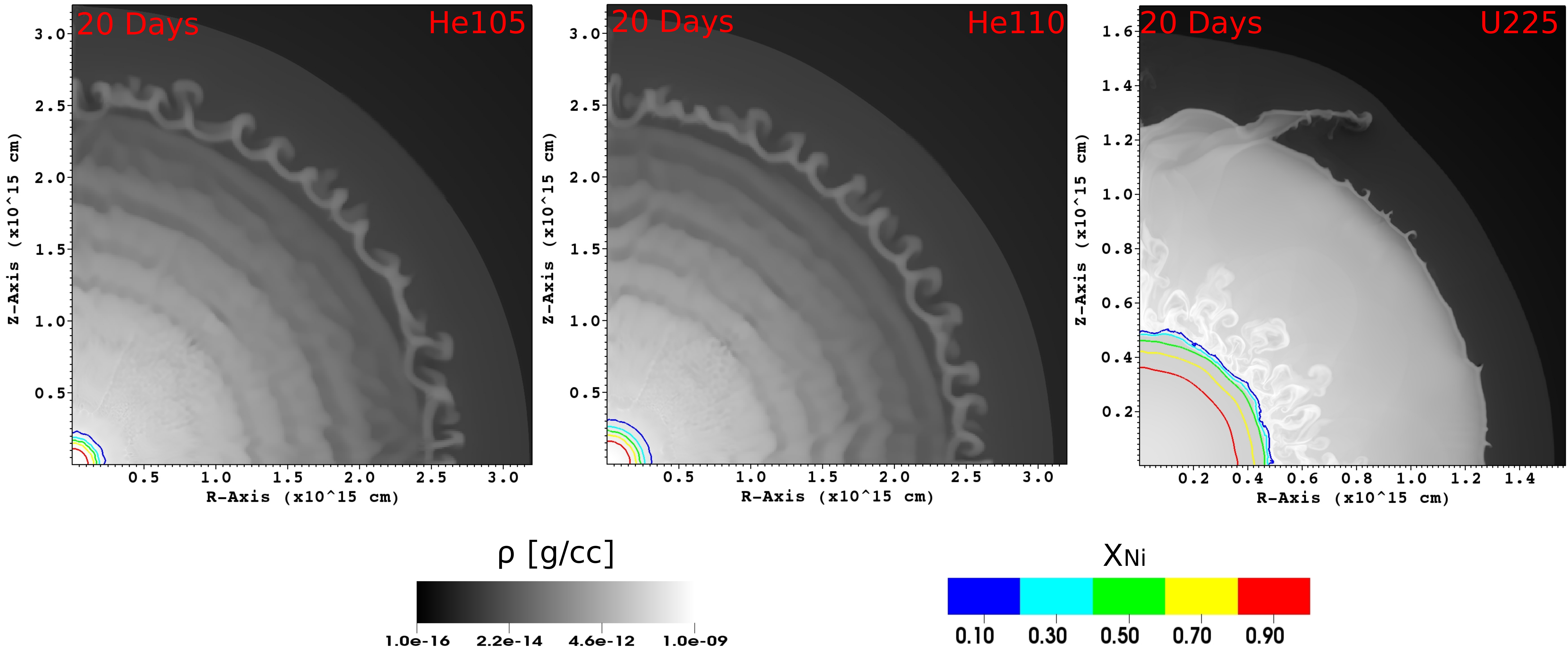} 
	\includegraphics[width=\textwidth]{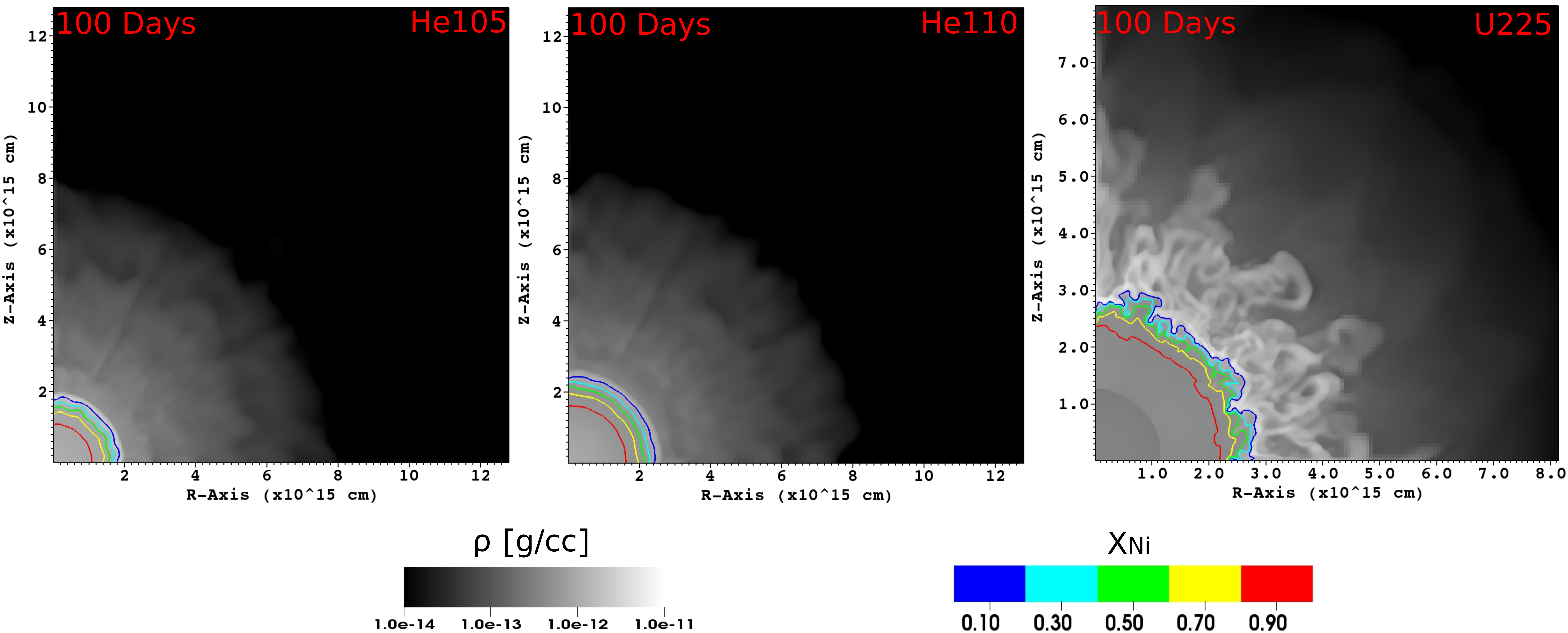} 
	\includegraphics[width=\textwidth]{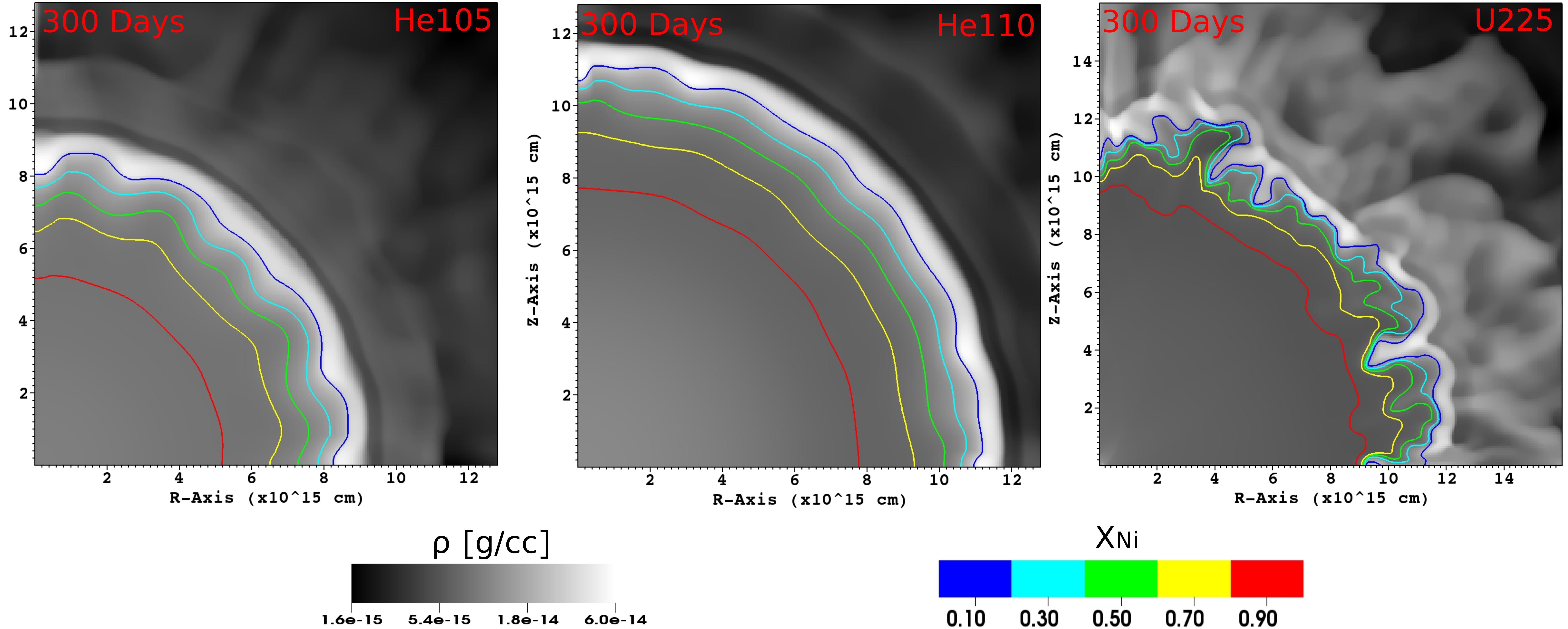} 
	\caption{\lFig{fig:evolution} Densities and \Ni\ contours in He105, He110, and U225 at 20, 100, and 300 days.  They remain mostly spherical in He105 and He110 but are disrupted by RT instabilities in U225, which dredge up some \Ni\ by 100 days.  A dense shell plowed up by the expansion of the hot \Ni\ bubble is clearly visible in all three models.  No mixing due to the expansion of the bubble occurs in any of the models (the mixing in U225 is due to instabilities formed by the reverse shock, not the expansion of \Ni).}
\end{figure*}

\begin{figure*}
	\centering
	\includegraphics[width=.8\textwidth]{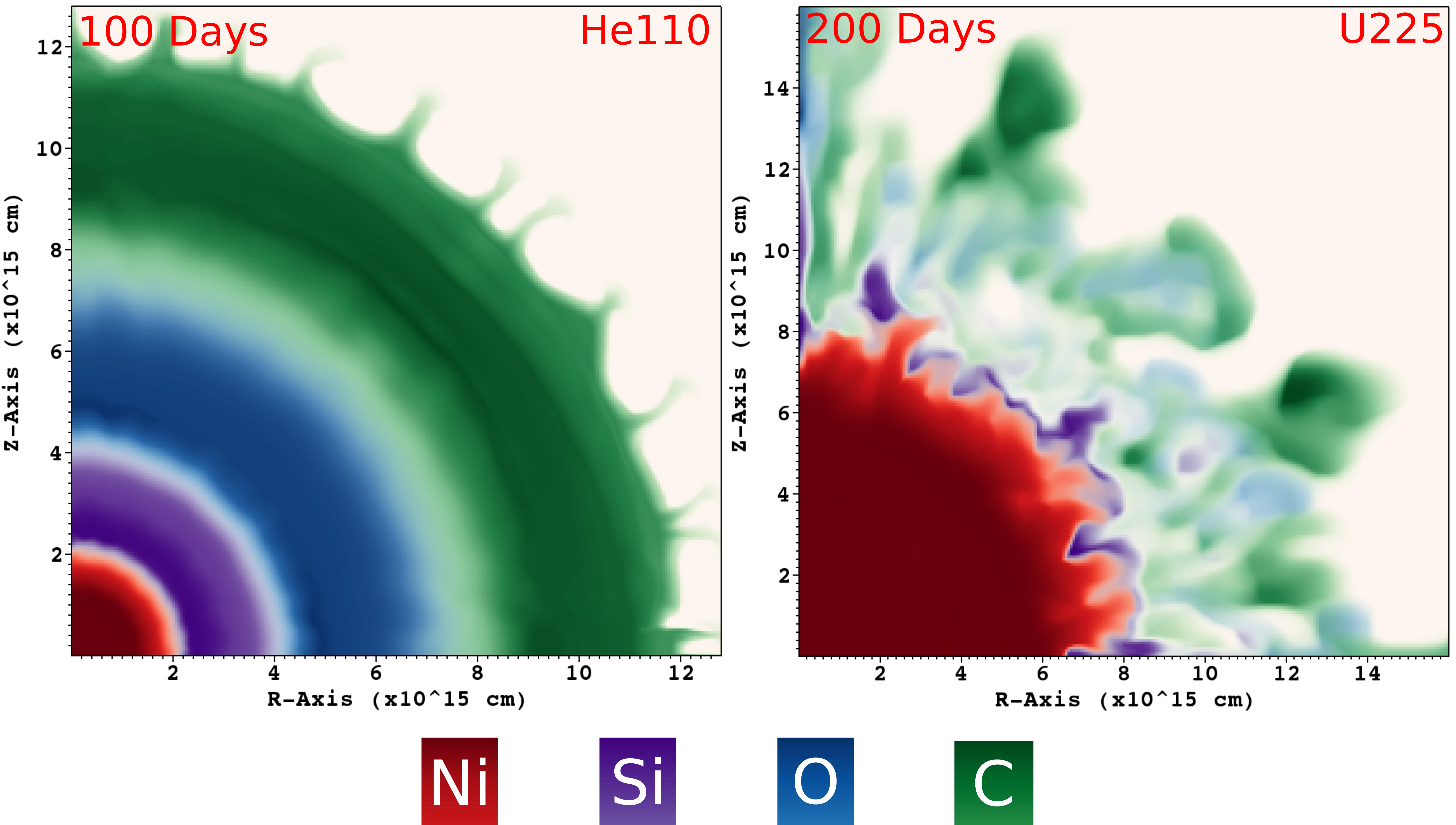} 
	\caption{\lFig{fig:mixing} Mixing in the \Cx\ (green), \Ox\ (blue), \Si\ (purple) and \Ni\ (red) layers at 100 days in He110 and 200 days in U225, when the instabilities are most prominent.  With the exception of \Cx, no mixing has occurred in He110 and the \Ni\ in U225 remains largely intact in spite of the disruption of the shells around it.}
\end{figure*}

We plot mass fractions, energy release rates, and total energy released for one \Ms\ of \Ni\ in \Fig{fig:nidecay}.  A total of $1.8\times10^{50}$ erg is released from its decay over 250 days, $5.92\times10^{49}$ erg from $\Ni\rightarrow\Co$ and $ 1.28\times10^{50}$ erg from $\Co\rightarrow \Fe$.  Assuming that all the decay energy exits the ejecta as radiation at later times (100 days after explosion), $\geq 5$ \Ms\ \Ni\ is required for a PI SN become a superluminous supernova (SLSN) with a total radiation energy budget $\gtrsim$ $10^{51}$ erg.  

\section{Ejecta Dynamics}

Gas densities and \Ni\ mass fractions for all three explosions are shown at 20, 100, and 300 days in \Fig{fig:evolution}.  Since we do not trace the advection of \Co\ and \Fe\ they are included in the \Ni\ (their mass fractions can be extracted from \Fig{fig:nidecay}).  All three stars are completely destroyed, with no compact remnants left behind.  Most of the energy from \Ni\ decay has been released by 20 days, and Rayleigh-Taylor (RT) fingers have appeared behind the forward shock in the He105 and He110 models.  They are caused by a reflection wave rebounding inward from the surface of the star when the forward shock breaks through it.  Their amplitudes remain small and never reach the \Ni\ at the center.
  
More prominent RT instabilities appear in the U225 explosion that are due to the formation of a reverse shock when the forward shock begins to plow up the H layer of the star.  When the shock enters this extended envelope it decelerates and a reverse shock forms and then detaches from it.  The two shocks become separated by a contact discontinuity that is prone to RT instabilities if 
 \begin{equation}
 \frac{\partial \rho}{\partial r}\frac{\partial P}{\partial r} < 0,
 \end{equation}
where $\rho$ is the gas density and $P$ is the gas pressure.  Since the density of the ejecta  decreases with radius, the reverse shock creates a pressure inversion that is in the opposite direction of its density gradient.  The contact discontinuity destabilizes and RT fingers appear \citep[such features have been found in earlier simulations of RSG explosions;][]{chen14a}.  They form much closer to the \Ni-rich core than in the He105 and He110 runs and have already begun to perturb its outer layers. 

\Co\ becomes the dominant source of decay energy 60 days after the explosion and small RT fingers appear in the carbon shell at 100 days in He105 and He110 but they do not affect the \Ni, as shown in \Fig{fig:mixing}.  In contrast, the RT instabilities visible at 20 days in the U225 model have now disrupted the outer \Ni\ layers.  At 300 days more than $95\%$ of all the decay energy has been released and the \Ni\ bubble has grown to $r \sim 10^{16}$ cm in all three SNe.  Heat due to radioactive decay has caused the bubble to expand into surrounding the surrounding ejecta, plowing it up into a shell.  The thickness of the shell, $\delta r $, is $\sim 10^{15}$ cm and it remains mostly spherical in He105 and He110 but has become heavily disrupted in U225.     

\begin{figure*}[h]
	\centering
	\includegraphics[width=1.\columnwidth]{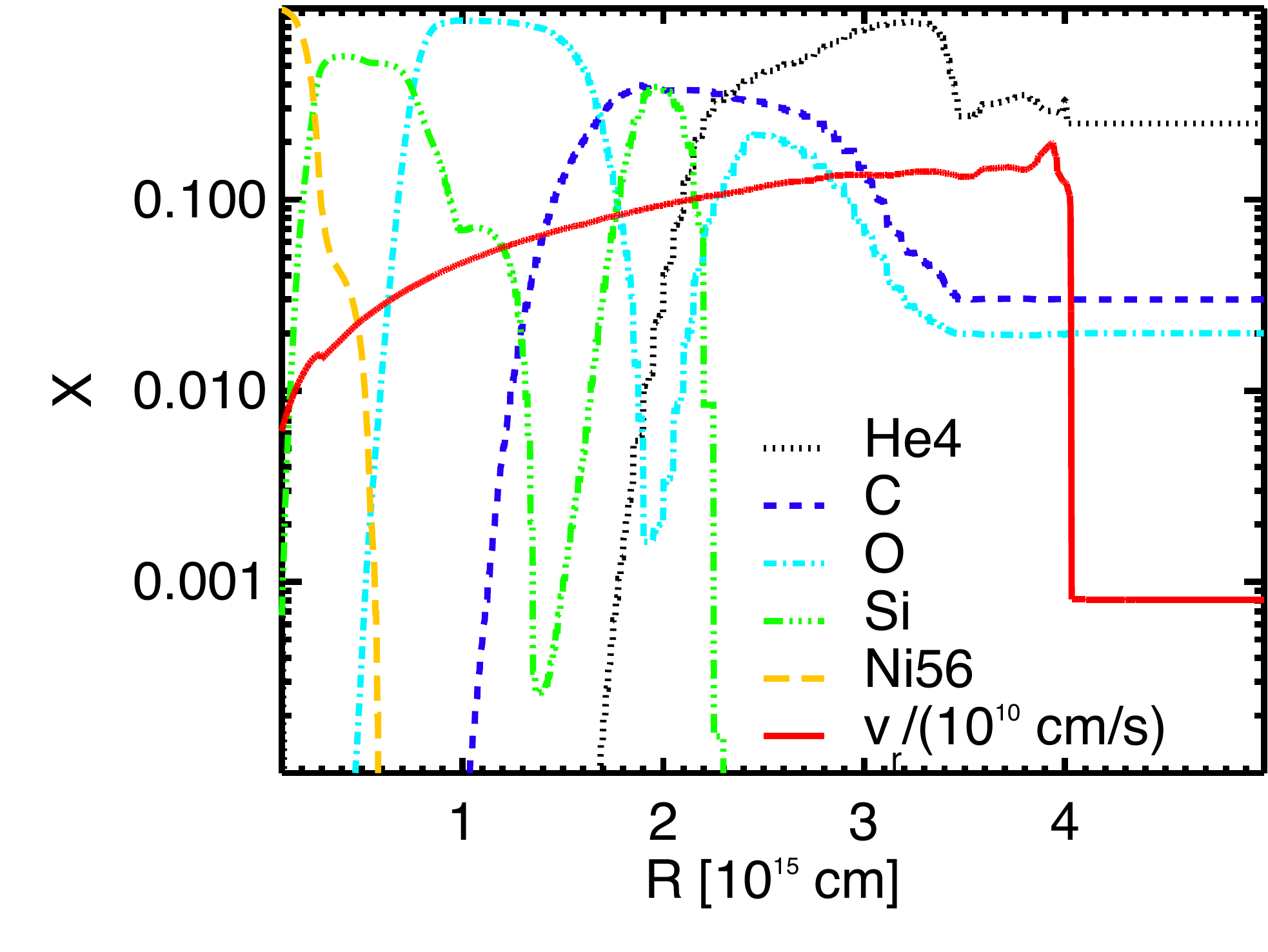} 
		\includegraphics[width=1.\columnwidth]{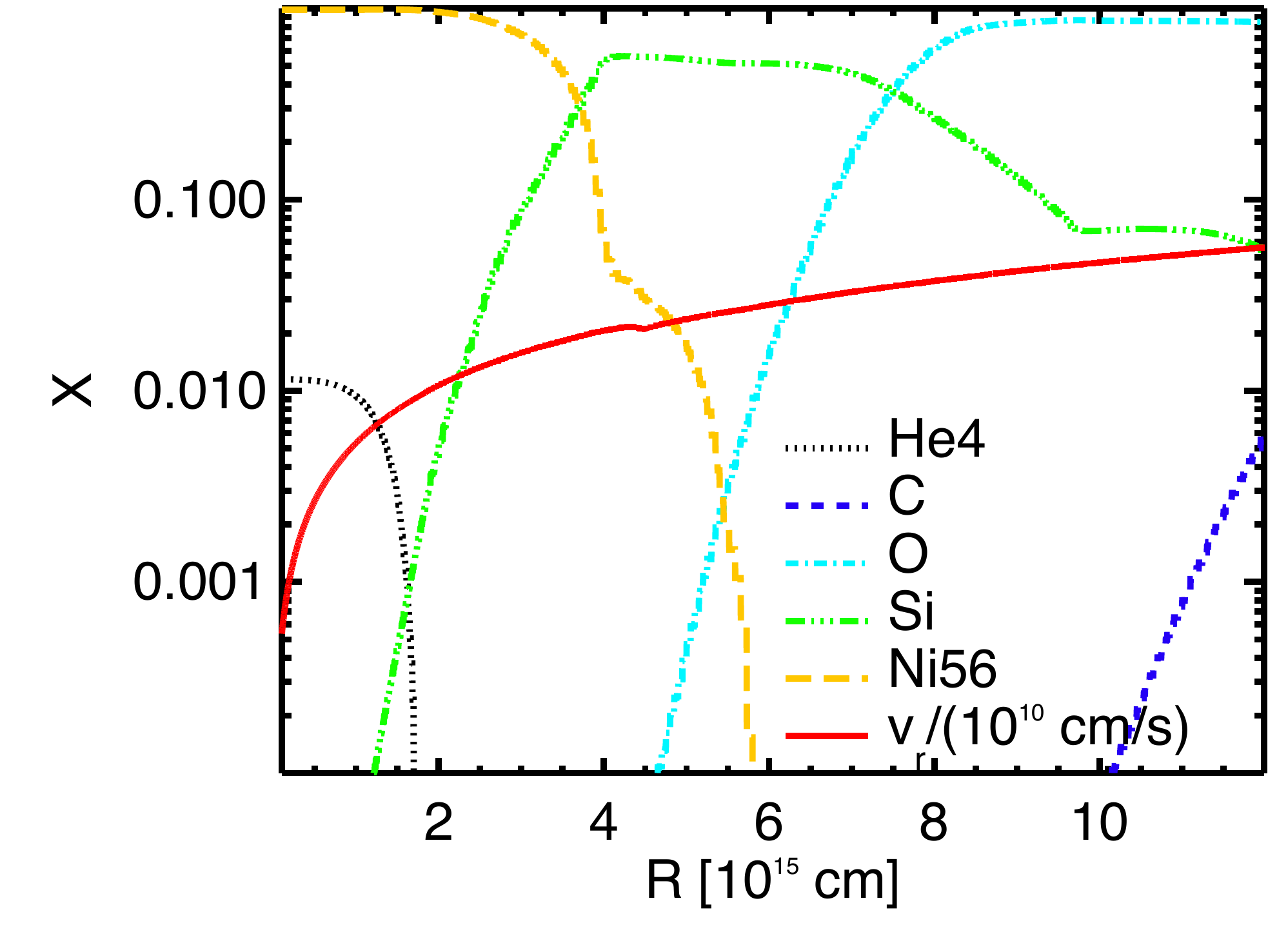} 
	\includegraphics[width=1.\columnwidth]{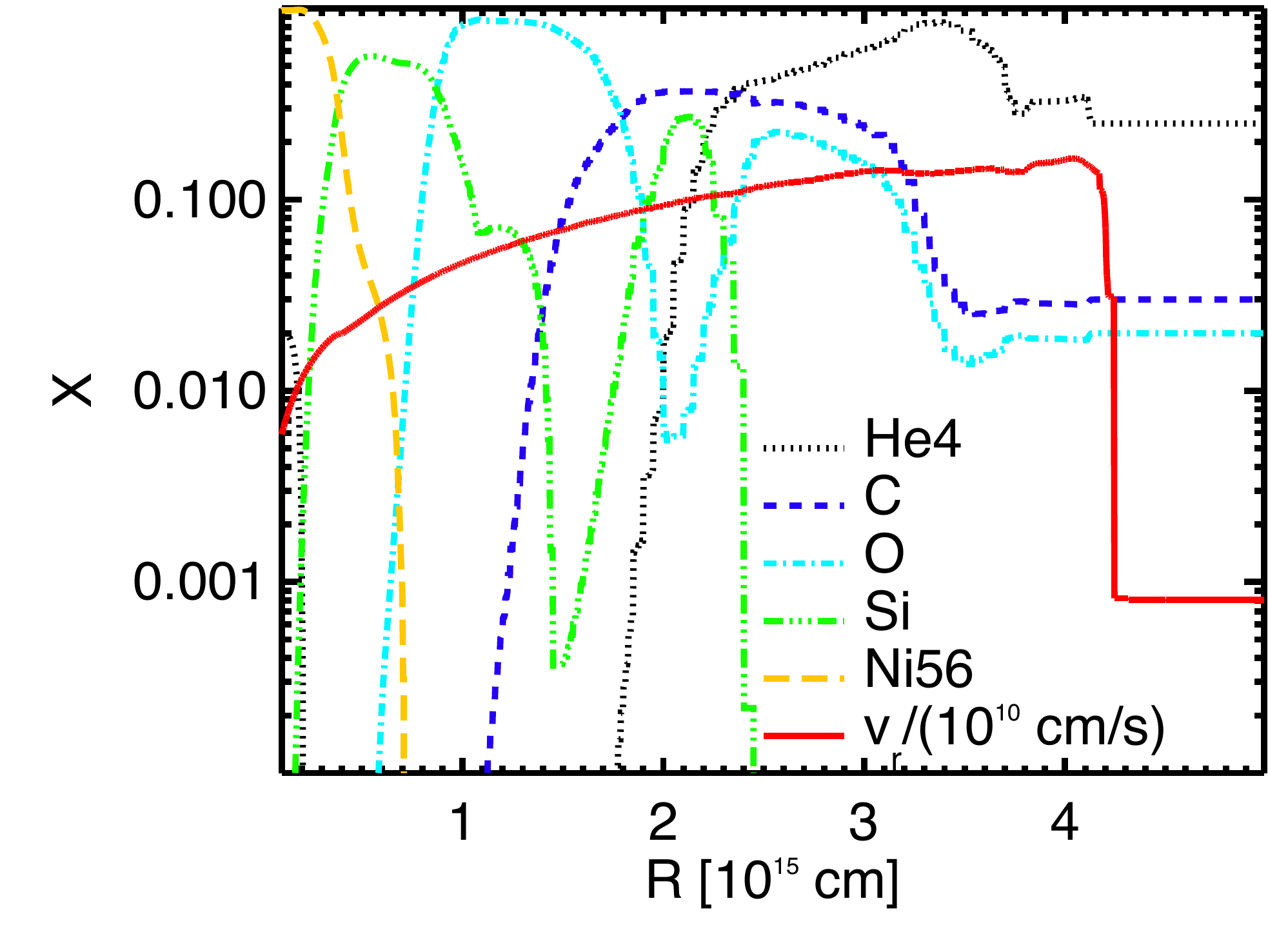}
		\includegraphics[width=1.\columnwidth]{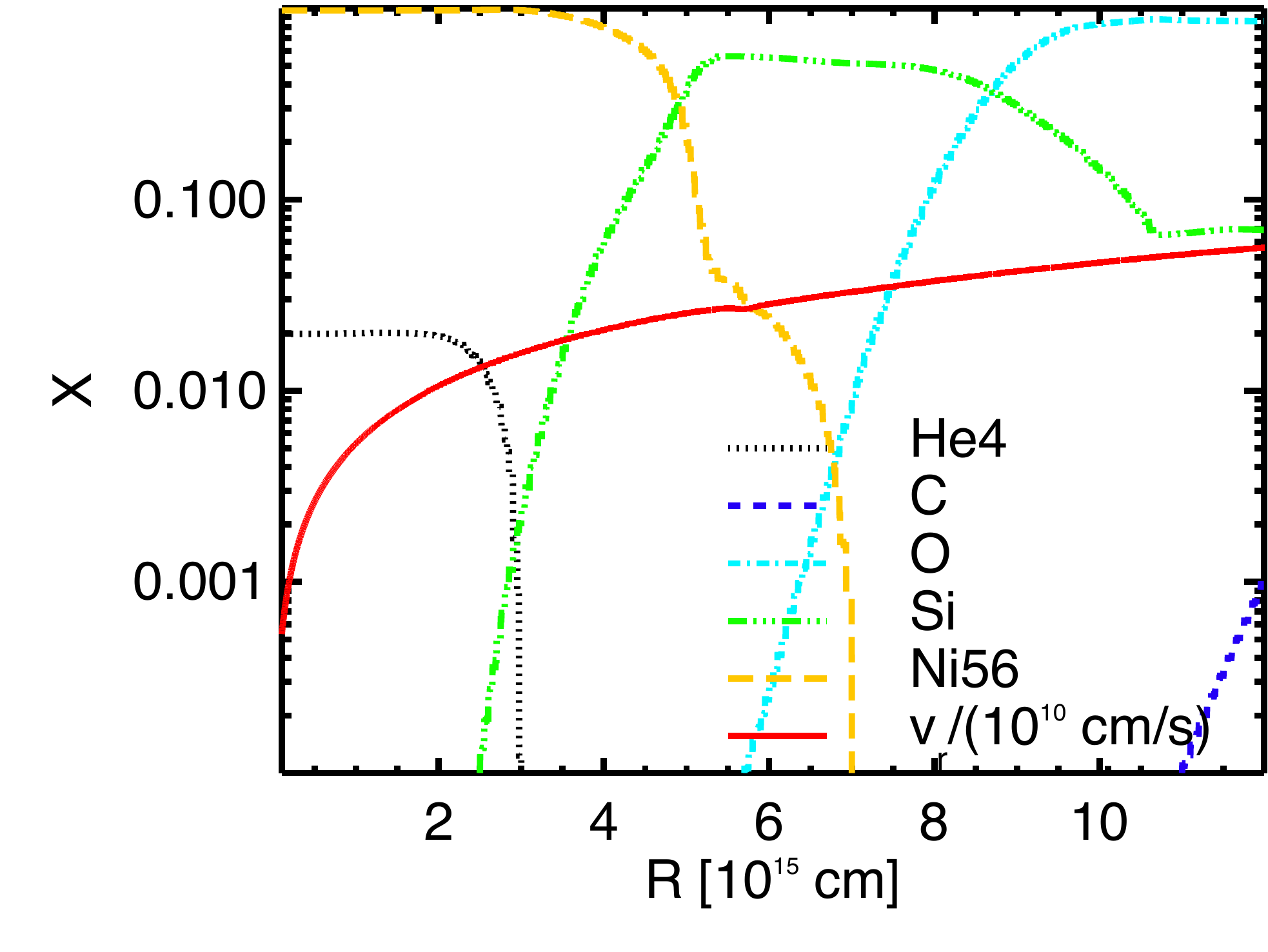}
	\includegraphics[width=1.\columnwidth]{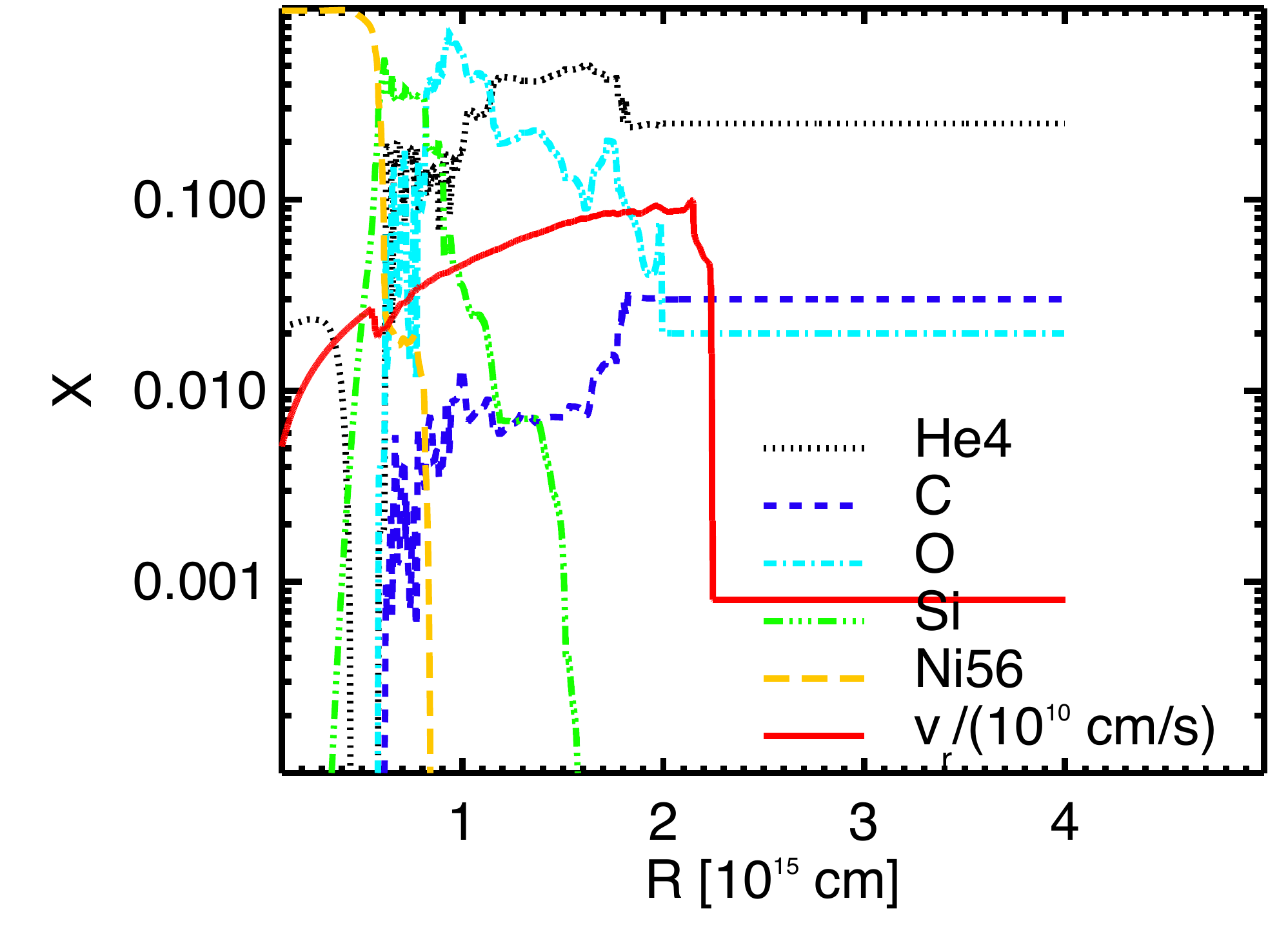}  
	\includegraphics[width=1.\columnwidth]{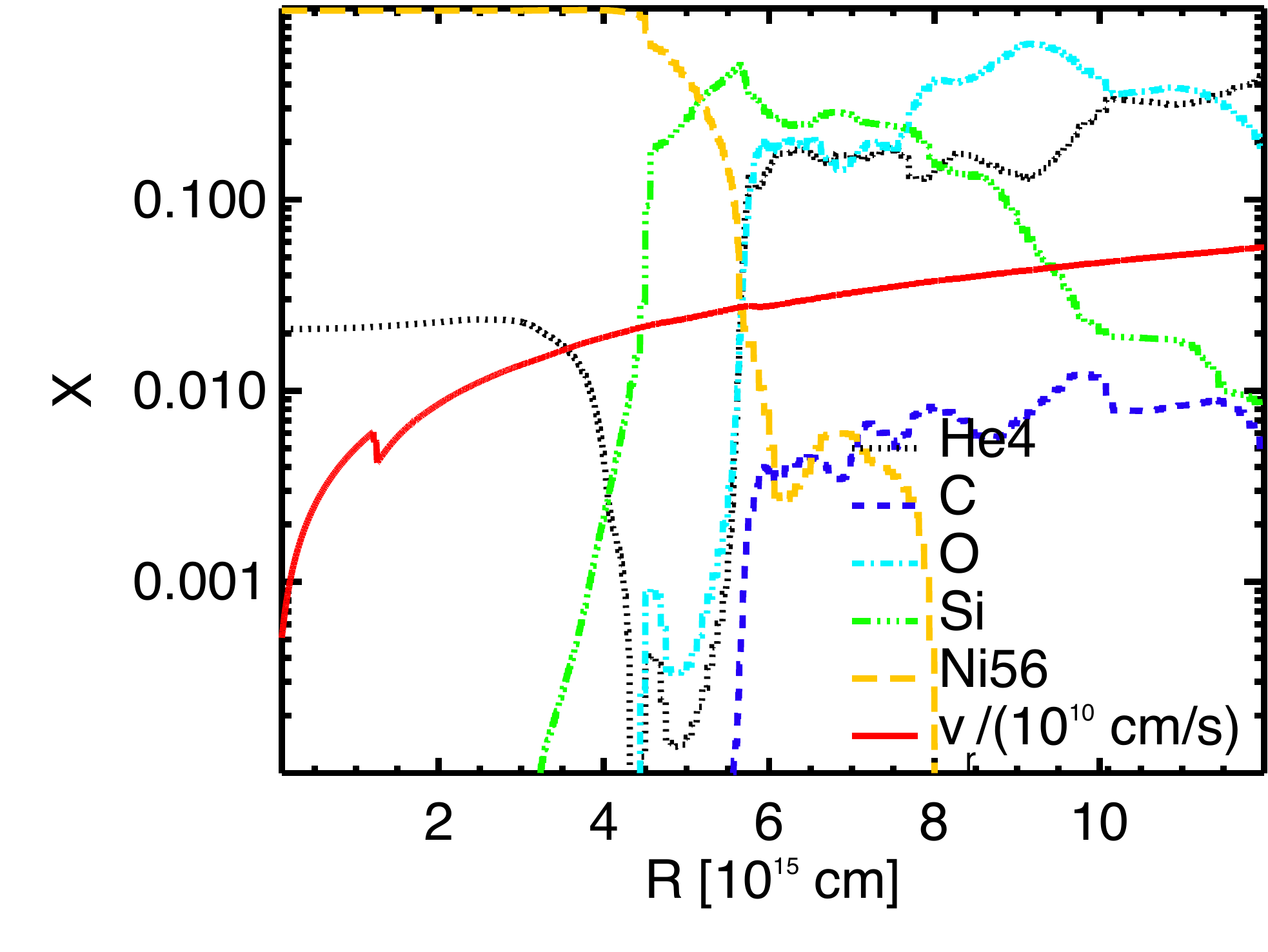}  			
	\caption{\lFig{fig:elements}  \He, \Cx, \Ox, \Si\ and \Ni\ mass fractions at 20 (left panels) and 200 days (right panels) after explosion.  Top:  He105; center:  He110; bottom:  U225.  Velocity profiles are superimposed on these mass fractions for comparison.  Mixing in U225 is clearly visible as the distortion in abundances at later times.}
\end{figure*}

We show the \Ni, \Si, \Ox\ and \Cx\ layers in the He110 and U225 runs at 100 and 200 days in  \Fig{fig:mixing}.  \ken{These times capture instabilities and mixing when they are most prominent in these regions.  Trace amounts of mixing appear in \Cx\ in He110 but not in the other three layers, which remain essentially frozen in mass coordinate out to 300 days.  The small RT fingers in the outer layer of \Cx\ form when a weak reverse shock steps back through the ejecta as the SN shock plows up the CSM.  It is weak because the CSM is diffuse.  Mixing is much more extensive in U225, with severe disruptions of the \Cx, \Ox\ and \Si\ shells and the outer layers of the \Ni\ core caused by instabilities driven by the reverse shock.}  
   
\subsection{Mixing Due to \Ni\ Heating}

To better quantify mixing in our runs we plot 1D angle-averaged mass fractions for \He, \Cx, \Ox, \Si\ and \Ni\ at 20 and 200 days after explosion in \Fig{fig:elements}.  At 20 days \Ni\ overlaps with \Si\ in all three SNe because it forms from explosive \Si\ burning, but it also extends slightly into the lower layers of the \Ox\ shell.  In contrast, \Ni\ and \Cx\ never mix in the helium star explosions.  In the U225 run \Cx, \Ox, and \Si\ all appear in the outer layers of the \Ni\ core, indicating that RT instabilities have already caused mixing in the innermost regions of the ejecta by 20 days.  The velocity of the forward shock at this time is $1 - 2\times 10^9$ cm s$^{-1}$ in all three models.  The \He\ at the center of the ejecta is due to photodisintegration of \Ni\ during explosive burning.  \Ni\ is formed during the first 20 s of the explosion in \Si\ but then begins to be destroyed at 15 - 20 s by thermal photons created in the extreme temperatures of rapid burning \citep[see Figure 5 of][]{chen14a}.  Note that the CSM of the two helium stars has H, \He, \Cx\ and \Ox\ mass fractions of 0.7, 0.25, 0.03 and 0.02, consistent with the stripping away of the H envelope prior to explosion (these mass fractions were also applied to U225 for consistency).

By 200 days the forward shock has reached $r \sim 1.2 \times 10^{16}$ cm, $\sim 90\%$ of the energy due to radioactive decay has been released, and elemental abundances have begun to freeze in mass coordinate.  The \Ni\ and \Ox\ mass fractions still cross each other at $10^{-2}$, indicating that little \Ni\ has been dredged up in He105 and He110.  \Ni\ overlaps with \Ox\ and \Cx\ at $r \sim 6-8 \times 10^{15}$ cm in U225, but only small amounts of it appear in \Cx.  We show angle-averaged density profiles for all three models at 200 days in \Fig{fig:shell}.  The dense shell plowed up by the expansion of the hot \Ni\ bubble within the ejecta is at $r \sim 4-6 \times 10^{15}$ cm.  It has a width $\Delta r \sim 10^{15}$ cm and a density that is 3 - 4 times higher than its surroundings.  The absence of instabilities in the shell in the helium star runs at this time indicates that \Ni\ heating does not cause much mixing (the disruption of the shell in U225 is due to instabilities driven by the reverse shock, not radioactive decay).

Mixing at these depths in the ejecta could be greater in actual PI SNe than in our models because convection in the core of the star could produce hot spots, off-center ignitions and asymmetric eruptions that cannot occur in our 1D burn calculations.  These can cause prompt \Ni\ mixing and produce chemical yields that are different from those in our simulations.  Our models also do not include radiation transport so they cannot capture the decoupling of radiation and gas at later times, which is subject to a variety of radiation-hydrodynamical instabilities. 

The total \Ni\ decay energy in our models is $1.58 - 3.07\times10^{51}$ erg, about twenty times smaller than the explosion energy, so it plays only a minor role in the dynamics of the flows.  If all the decay energy is converted into kinetic energy at the base of the 40 - 50 \Ms\ \Si\ shell its net velocity gain would be $2 - 3 \times10^8$ cm s$^{-1}$, which is consistent with our simulations.  We find that this energy is only sufficient to plow up a shell, not accelerate it to velocities capable of driving the formation of fluid instabilities.   

\begin{figure}
	\centering
	\includegraphics[width=\columnwidth]{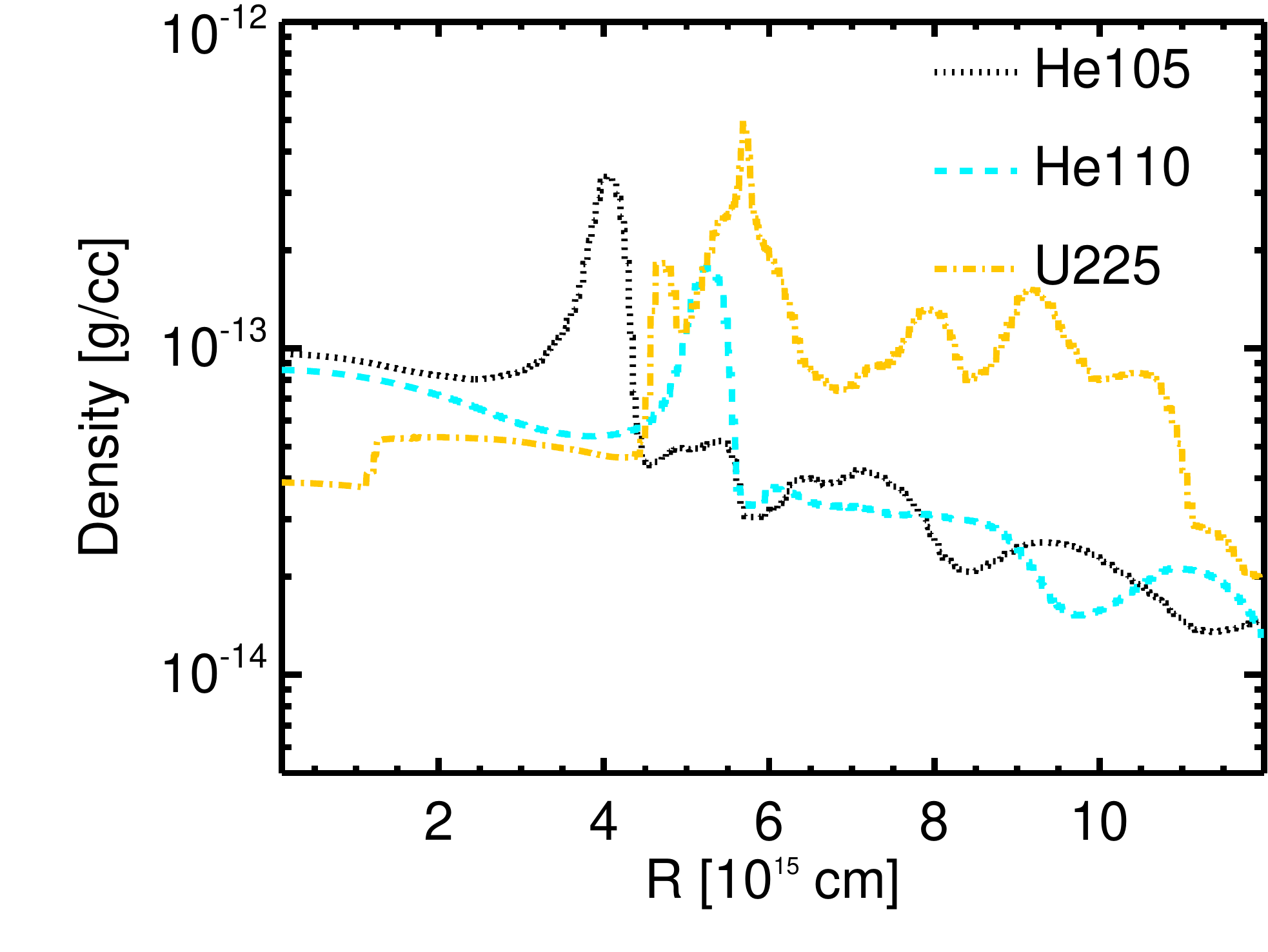} 
	\caption{\lFig{fig:shell} Density profile of the inner ejecta at 200 days.  The dense shells at $r \sim 4-6 \times10^{15}$ cm are plowed up by the expansion of the hot \Ni\ bubble.  The shell in U225 is partially broken up by RT instabilities driven by the reverse shock.}
\end{figure}

\section{Discussion}

\subsection{Comparison to Previous Models}

\ken{\citet{candace2011} performed the first multidimensional simulations of PI SNe and found little mixing but only evolved them for short times, ending them just after shock breakout.  \citet{chen2011} later studied both core contraction and explosion in 2D PI SN models and saw minor mixing in the oxygen burning shell but they too only evolved the blast for short times.  \citet{chat2012a} and \citet{chen14a} performed a suite of 2D PI SN runs with nuclear burning and, like \citet{candace2011}, found that mixing depends heavily on the structure of the progenitor and is stronger in RSGs than in BSGs.  Most recently, \citet{Gil17} carried out the first 3D PI SN simulations with nuclear burning and found minor mixing that was consistent with earlier work.  The longest that any of these studies evolved the explosion was less than a month \citep{chen14a} so they could not evaluate the effects of \Ni\ decay, but our runs exhibit similar degrees of mixing out to the times these earlier models were run.}

\subsection{Light Curves / Spectra}

\ken{Mixing deep in the ejecta could alter the spectra of PI SNe at late times by changing the order in which some lines appear.  Prominent \Si, \Mg, and \Ox\ lines appear at different times with maximum intensities that depend on mixing at the \Si\ / \Ox\ interface.  However, \citet{jer16} and \citet{ch19} recently calculated light curves and spectra from multidimensional PI SN models and found that mixing does not affect PI SN light curves and that their spectra and color evolution does not change with viewing angle.  However, the degree to which the expansion of the hot \Ni\ bubble diverts energy from rebrightening at later times remains unknown, but it will clearly produce transients that are less luminous than those predicted by most light curve calculations. }

\subsection{PI SNe as SLSNe}

If even half of the energy of the \Ni\ decay in our models escapes the ejecta as radiation it could easily power a SLSN with a peak bolometric luminosity that exceeds $10^{44}$ erg for 100 days.  If so, how could it be distinguished from magnetar-powered SNe, which have also been proposed as SLSNe?  The latter are powered by the spin-down energy of millisecond pulsars with extreme magnetic fields \citep{blondin2001,che11,chen16,chen17c,chen20a}.  Although both PI and magnetar-powered SNe can reproduce SLSN light curves, their central engines can be distinguished by their spectra.  Unlike PI SNe, mixing deep in the ejecta of magnetar-powered explosions is rampant and produces rapidly evolving spectra \citep{chen16}.  In contrast, the decay of large masses of \Ni\ should produce strong \Fe\ lines in the nebular phases of PI SNe. 

\subsection{Rotation}

Stellar rotation, although not included in our models, would likely reduce \Ni\ production, decay heating, hot bubble expansion, and therefore mixing deep in the ejecta.  As mentioned earlier, rotation can cause VMS to encounter the PI at lower masses because rotational mixing over the life of the star builds up more massive He cores for a given progenitor mass.  However, \citet{chen2015} found that centrifugal forces due to rotation can partially counteract core contraction and decrease explosion energies and \Ni\ synthesis in PI SNe.  Consequently, the PI SNe of rotating stars would exhibit less mixing at early times and less radioactive heating and internal expansion of the ejecta at later times.  

\subsection{Mixing in 3D}

How might mixing change in our models in 3D?  Numerical simulations by \citet{chen17} suggest that mixing is weaker in 3D because of the nature of turbulence.  Fluid instabilities are more violent in 2D than in 3D in a given model because turbulent cascades are stronger, so they produce more mixing.  Indeed, simulations are often run in 2D before being run at much higher cost in 3D to determine the upper limit to the mixing that would be expected in the problem.  \citet{Gil17} found that differences in mixing in 2D and 3D were most pronounced at the Si / O interface in PI SNe.  The RT fingers were similar but their overdensities were larger in 3D.  However, mixing is sensitive to grid artifacts and initial perturbations so it is not yet clear if mixing in PI SNe is stronger in 2D or 3D. 

\section{Conclusion}

We find that \Ni\ bubble dynamics does not affect the spectra of PI SNe but can reduce bolometric luminosities during rebrightening.  How \Ni\ decay energy is partitioned between the internal and kinetic energies of the ejecta and the radiation that escapes the ejecta in PI SNe remains uncertain.  Previous calculations of PI SN light curves often ignored the transformation of decay energy into ejecta dynamics.   \citet{Koz15} and \citet{kz17} calculated 1D PI SN light curves with \STELLA, which includes radiation hydrodynamics, so they were able to track deviations from the homologous expansion assumed in some earlier calculations \citep[e.g.,][]{det12}.  They found density structures created by \Ni\ bubble expansion that were consistent with those in our models but few effects on rebrightening.  However, their models were limited to very low mass resolution ($\sim$ 150 lagrangian zones) that may have produced spurious radiation transport.  The formation of dense shells due to \Ni\ expansion in our models would clearly come at the expense of rebrightening at later times.  

Mixing driven by processes other than \Ni\ expansion can alter the spectra of some PI SNe but not others.  For example, RT instabilities driven by the formation of reverse shocks in RSG explosions can jumble together the \Cx, \Ox\ and \Si\ shells at early times and affect the order in which their spectral lines emerge at later times when photons due to \Ni\ decay finally diffuse out of the ejecta.  Such mixing does not occur in the explosions of bare He cores so there are few if any changes to the order in which spectral lines later appear.

We note that even if some of the energy of radioactive decay is diverted from rebrightening into the internal expansion of the \Ni\ bubble, it does not disqualify PI SNe as SLSN candidates.  Only $\sim$ 5 \Ms\ of \Ni\ is required to produce a superluminous event if most of the energy of decay escapes the ejecta as photons.  Our models produce 8.5 - 16.5 \Ms\ of \Ni, so even if the majority of the decay energy is lost to work enough could still escape to create an extremely bright transient.  We are now developing high-resolution 1D radiation hydrodynamical simulations of PI SNe with \CASTRO\ with $\gamma$-ray transport to determine how energy due to \Ni\ decay is partitioned in PI SNe and its effects on rebrightening.

\acknowledgments

The authors thank the referee for their constructive comments and Alexander Heger for many useful discussions.  KC acknowledges support from an EACOA Fellowship and Ministry of Science and Technology (Taiwan, R.O.C.) grant number MOST 107-2112-M-001-044-MY3.  KC also thanks the hospitality of the Aspen Center for Physics, which is supported by NSF grant PHY-1066293, and the Kavli Institute for Theoretical Physics, which is supported by NSF grant PHY-1748958, the Nordic Institute for Theoretical Physics (NORDITA), and Physics Division at the National Center for Theoretical Sciences (NCTS).  D. J. W. was supported by the Ida Pfeiffer Professorship at the Institute of Astrophysics at the University of Vienna.   Our numerical simulations were done at the National Energy Research Scientific Computing Center (NERSC), a U.S. Department of Energy Office of Science User Facility operated under Contract No. DE-AC02-05CH11231, the Center for Computational Astrophysics (CfCA) at the National Astronomical Observatory of Japan (NAOJ), and the TIARA Cluster at the Academia Sinica Institute of Astronomy and Astrophysics (ASIAA).

\software{\CASTRO\  \citep{ann2010, zhang2011}},  \KEPLER\ \citep{kepler},


\end{document}